\documentclass[a4paper,11pt]{article}
\usepackage{jheppub} % for details on the use of the package, please see the JINST-author-manual

\pdfoutput=1

\usepackage{tikz}
    \usetikzlibrary{positioning}
    \usetikzlibrary{arrows}
    \usetikzlibrary{shapes}
    \usepackage{pgfplots}
    \usetikzlibrary{calc}
    \usetikzlibrary{decorations.markings}
     \usetikzlibrary{decorations.pathmorphing}
    \tikzset{snake it/.style={decorate, decoration=snake}}
\def\centerarc[#1](#2)(#3:#4:#5) % Syntax: [draw options] (center) (initial angle:final angle:radius)
    { \draw[#1] ($(#2)+({#5*cos(#3)},{#5*sin(#3)})$) arc (#3:#4:#5); }
    \usepackage{booktabs}
\usepackage{array}
\usepackage{jheppub}
\usepackage{amsmath}
\usepackage{amssymb}
\usepackage{amsfonts}
\usepackage{amsthm}
\usepackage{shuffle}
\usepackage{enumitem}
\usepackage{color}   
\usepackage{hyperref}
\usepackage{bbm}
\usepackage{multirow}
\usepackage{comment}
\usepackage{dsfont}
\newcommand{\ord}{\begin{cal}O\end{cal}}

\def\beq{\begin{equation}}
\def\eeq{\end{equation}}
\def\bsp#1\esp{\begin{split}#1\end{split}}

\newenvironment{sloppyequation}[0]{\sloppy\begin{flushleft}\hspace*{0.75cm}\(}{\)\end{flushleft}\fussy}

\newcommand{\beqsloppy}{\begin{sloppyequation}}
\newcommand{\eeqsloppy}{\end{sloppyequation}}

\newcommand{\cF}{\begin{cal}F\end{cal}}
\newcommand{\cG}{\begin{cal}G\end{cal}}

\newcommand{\cM}{\begin{cal}M\end{cal}}
\newcommand{\cN}{\begin{cal}N\end{cal}}

\newcommand{\bbI}{\mathbb{I}}

\usepackage{hyperref}
\theoremstyle{definition}

\newcommand{\uz}{{z}}

\newcommand{\uPi}{{\Pi}}

\DeclareMathOperator{\SL}{SL}
\DeclareMathOperator{\K}{K}

\definecolor{ourblue}{RGB}{0, 57, 120} % 
\colorlet{ourlightblue}{ourblue!50!white}
\newcommand{\bs}[1]{\boldsymbol{#1}}
\newcommand{\eps}{\varepsilon}
\newcommand{\rd}{\mathrm{d}}

\newcommand{\bx}{\bs{x}}

\newcommand{\bz}{\bs{z}}
\newcommand{\bA}{\bs{A}}

\newcommand{\bI}{\bs{I}}
\newcommand{\bJ}{\bs{J}}

\newcommand{\bM}{\bs{M}}

\newcommand{\bR}{\bs{R}}

\newcommand{\bOmega}{\bs{\Omega}}

\def\beq{\begin{equation}}
\def\eeq{\end{equation}}
\def\bsp#1\esp{\begin{split}#1\end{split}}

\allowdisplaybreaks

\title{Three-loop banana integrals with three equal masses}

\author[a]{Claude Duhr}
\emailAdd{cduhr@uni-bonn.de}
\author[a]{Sara Maggio}
\emailAdd{smaggio@uni-bonn.de}
\affiliation[a]{Bethe Center for Theoretical Physics, Universit\"at Bonn, D-53115, Germany
}

\abstract{We obtain and solve the canonical differential equations for the three-loop banana integrals in dimensional regularisation when three of the four masses are equal. The K3 surface associated with the maximal cuts factorises into a product of two elliptic curves. This allows us to express the differential forms in the canonical differential equations in terms of meromorphic modular forms. We present a rigorous proof that these differential forms  only have simple poles and that they define independent cohomology classes. We also present explicit results for all master integrals in terms of iterated integrals of meromorphic modular forms (and integrals thereof). This is the first time that it was possible to express the results of a Feynman integral associated with a K3 geometry and depending on two dimensionless ratios in terms of functions that have previously been studied in the literature.}

\begin{document}

\preprint{BONN-TH-2025-32}

\maketitle

\section{Introduction}
\label{sec:introduction}

Over the last decade, a lot of progress has been made in analytically evaluating dimensionally-regulated Feynman integrals and in understanding the relevant class of transcendental functions. During this process it has become evident that these functions and their properties are tightly linked to periods of algebraic varieties~\cite{MR1852188,Bogner:2007mn} and iterated integrals~\cite{ChenSymbol}. For example, multiple polylogarithms (MPLs) (cf.,~e.g.,~refs.~\cite{Mpls1,Remiddi:1999ew,Gehrmann:2000zt,Ablinger:2011te}) are closely related to the geometry of the punctured Riemann sphere. In cases where the underlying geometry involves an elliptic curve, the natural class of functions are  elliptic multiple polylogarithms (eMPLs)~\cite{MR1265553,LevinRacinet,BrownLevin,Broedel:2014vla,Broedel:2017kkb,EnriquezZerbini} and iterated integrals of modular forms~\cite{ManinModular,Brown:2014pnb,Adams:2017ejb,Matthes2022IteratedPrimitives,Broedel:2021zij}. Generalisations of polylogarithms to higher-genus Riemann surfaces have also recently been defined~\cite{DHoker:2023vax,DHoker:2024ozn,DHoker:2025szl,Baune:2024ber,Baune:2024biq}.

The aforementioned classes of special functions correspond to underlying geometries of complex dimension one. Also higher-dimensional complex manifolds can arise from Feynman integral computations. The most prominent examples are Calabi-Yau (CY) varieties (see for example ref.~\cite{Bourjaily:2022bwx} for a review where they appear). In particular, the maximal cuts of $L$-loop banana integrals compute the periods of a family of CY $(L-1)$-folds~\cite{Bloch:2014qca,MR3780269,Bonisch:2020qmm,Klemm:2019dbm,Bonisch:2021yfw}.\footnote{See also refs.~\cite{Kreimer:2022fxm,Mishnyakov:2023sly,Mishnyakov:2023wpd,Mishnyakov:2024rmb}.} However, the resulting class of iterated integrals is in general not understood at the same level as for the aforementioned one-dimensional geometries. It is an interesting question if there are special mass configurations for which the banana integrals can be expressed in terms of transcendental functions that have already been introduced and studied in the mathematics literature. For example, the geometry underlying the two-loop sunrise integral is a family of elliptic curves, and the relevant class of functions can be identified with eMPLs and iterated integrals of modular forms~\cite{Bloch:2013tra,Broedel:2017siw,Adams:2017ejb,Bogner:2019lfa} (see also refs.~\cite{Adams:2013nia,Adams:2014vja,Adams:2015ydq}). At higher loops, this question is closely related to the situation when the underlying CY manifold reduces to a simpler geometry. 
This is particularly well understood for CY twofolds (also known as K3 surfaces)~\cite{doran,BognerCY,BognerThesis,10.1215/ijm/1258138437,Almkvist3,Clingher2010LatticePK,doranclingher,doranclingher1}. For example, it is well known that the maximal cuts of the three-loop equal-mass banana integral can be written as the symmetric square of the periods of the sunrise elliptic curve~\cite{Joyce1973SimpleCubicLGF,Bloch:2014qca,Primo:2017ipr}. This result was used in refs.~\cite{Broedel:2019kmn,Broedel:2021zij,Pogel:2022yat} to express the equal-mass three-loop banana integrals in terms of iterated integrals of (meromorphic) modular forms. 

Recently, it was shown that also the periods of the K3 surface attached to the three-loop banana integral where three of the four masses are equal can be written as products of modular forms~\cite{Duhr:2025tdf,Duhr:2025ppd}. 
%While this implies that also for the case of three-equal mass a representation in terms of iterated integrals of modular forms should exist, so far this result has not been presented in the literature.
%
The main goal of this paper is to present for the first time fully analytic results for these integrals in terms of iterated integrals of meromorphic modular forms (and integrals thereof). We use the method of differential equations~\cite{Kotikov:1990kg,Kotikov:1991hm,Kotikov:1991pm,Remiddi:1997ny,Gehrmann:1999as}, and our main strategy consists in transforming the differential equations into a canonical form~\cite{Henn:2013pwa}. While there is currently no general consensus what the correct definition of a canonical form is in the presence of general geometries, a lot of progress has recently been made in the case of (hyper)elliptic or CY manifolds~\cite{Gorges:2023zgv,Pogel:2022ken,Pogel:2022vat,Pogel:2022yat,Duhr:2024uid,Duhr:2025lbz,Maggio:2025jel,e-collaboration:2025frv,Bree:2025tug} (see also refs.~\cite{Chen:2025hzq,Chaubey:2025adn}). In particular, very recently a canonical basis for the three-loop banana integrals with four distinct masses was obtained~\cite{Duhr:2025kkq,Pogel:2025bca}. The solution to these canonical differential equations involves iterated integrals over kernels that contain the periods of the underlying family of K3 surfaces, and it is expected that these iterated integrals define new classes of transcendental functions. For the case of three equal masses, we can go further: by using the fact that the periods can be cast in the form of a product of modular forms, we can choose a path of integration on which all iterated integrals can be evaluated in terms of iterated integrals of modular forms (and integrals of the latter). This is the first time that it was possible to express the result of a Feynman integral involving a family of CY varieties depending on two (dimensionless) variables in terms of transcendental functions that had already been defined in the literature. We can then use known results for iterated integrals of modular forms to understand their properties. As an example, we show that all the differential forms that appear in the canonical differential equations are independent and only have simple pole, which are properties that canonical differential equations are conjectured to satisfy~\cite{Duhr:2025lbz}.

Our paper is organised as follows. In section~\ref{sec:banana} we review known results about (the maximal cuts of) the three-loop banana integrals with three or four equal masses. In section~\ref{sec:canonicaldeq} we introduce the canonical differential equations for the three-loop banana integrals with three equal masses and we present the main results of our paper, namely the analytic results for the integrals in terms of iterated integrals of modular forms. In section~\ref{sec:conclusions} we draw our conclusions. We include appendices where we review modular forms and where we collect the explicit expressions for the differential forms that appear in the canonical differential equations. There, we also present some proofs omitted in the main text. Together with the arXiv submission, we provide ancillary files containing the canonical differential equations, the differential forms and the solution in terms of iterated integrals.

% !TEX root = main.tex

\section{Three-loop banana integral with three equal masses}
\label{sec:banana}
%-----Picture 3-Loop Banana Graph-------------
\begin{figure}[!th]
\centering
\begin{tikzpicture}
\coordinate (llinks) at (-2.5,0);
\coordinate (rrechts) at (2.5,0);
\coordinate (links) at (-1.5,0);
\coordinate (rechts) at (1.5,0);
\begin{scope}[very thick,decoration={
    markings,
    mark=at position 0.5 with {\arrow{>}}}
    ] 
\draw [-, thick,postaction={decorate}] (links) to [bend right=25]  (rechts);
\draw [-, thick,postaction={decorate}] (links) to [bend right=-25]  (rechts);

\draw [-, thick,postaction={decorate}] (links) to [bend left=85]  (rechts);
\draw [-, thick,postaction={decorate}] (llinks) to [bend right=0]  (links);
\draw [-, thick,postaction={decorate}] (rechts) to [bend right=0]  (rrechts);
\end{scope}
\begin{scope}[very thick,decoration={
    markings,
    mark=at position 0.5 with {\arrow{>}}}
    ]
\draw [-, thick,postaction={decorate}] (links) to  [bend right=85] (rechts);
\end{scope}
\node (d1) at (0,1.1) [font=\scriptsize, text width=.2 cm]{$m_1$};
\node (d2) at (0,0.6) [font=\scriptsize, text width=.2 cm]{$m_1$};
\node (d3) at (0,-0.15) [font=\scriptsize, text width=.2 cm]{$m_1$};
\node (d4) at (0,-.7) [font=\scriptsize, text width=0.3 cm]{$m_2$};
\node (p1) at (-2.0,.25) [font=\scriptsize, text width=1 cm]{$p$};
\node (p2) at (2.4,.25) [font=\scriptsize, text width=1 cm]{$p$};
\end{tikzpicture}
\caption{The three-loop banana graph with three equal masses.}
\label{fig:banana}
\end{figure}
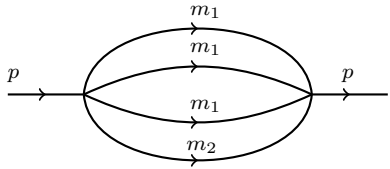
%------------------------
In this paper we consider three-loop banana integrals depending on four non-zero propagator masses, with three of the four masses being equal (see figure~\ref{fig:banana}). All integrals of this type are elements of the integral family defined by
\begin{align}
\label{eq:banana_family_def}
&I_{a_1,a_2,a_3,a_4,a_5,a_6,a_7,a_8,a_9} = \\
\nonumber  &= e^{3\gamma_E\eps} \int\! \Bigg(\prod_{j=1}^3\frac{\mathrm{d}^Dk_j}{i\pi^\frac{D}{2}}\Bigg)
\frac{ \, (k_3^2)^{-a_5} (k_1 \cdot p)^{-a_6} (k_2 \cdot p)^{-a_7} (k_3 \cdot p)^{-a_8} (k_1 \cdot k_2)^{-a_9}}
{[k_1^2 - m_1^2]^{a_1} [k_2^2 - m_2^2]^{a_2} [(k_1 - k_3)^2 - m_1^2]^{a_3} [(k_2 - k_3 - p)^2 - m_1^2]^{a_4}}\,.
\end{align} 
Here $\gamma_E=-\Gamma'(1)$ is the Euler-Mascheroni constant and we work in dimensional regularisation~\cite{tHooft:1972tcz,Cicuta:1972jf,Bollini:1972ui} in $D=2-2\eps$ dimensions, and the exponents $a_i$ of the propagators are integers.
It is easy to see that the only non-trivial functional dependence of the integrals can be through the ratios
\beq
z_i = \frac{m_i^2}{p^2}\,,\qquad i=1,2\,.
\eeq
We therefore set $p^2=1$ without loss of generality.

Not all integrals from this family are independent, but they are related by integration-by-parts (IBP) relations~\cite{Chetyrkin:1981qh,Tkachov:1981wb}. The IBP relations allow us to express every member of this integral family in terms of a finite number of so-called \emph{master integrals}. We solve the IBP relations using  \texttt{LiteRed2}~\cite{Lee:2013mka}, and we identify the following set of seven master integrals: 
%
%The maximal cuts compute the periods of a two-parameter family of K3 surfaces parametrised by $z_i=m_i^2/p^2$.
%
%The goal of this paper is to write the differential equations fully analytically in closed form for this three-loop integral. To do so, we use the results in Ref.~\cite{Duhr:2025ppd} to write the periods in terms of elliptic functions.
%
%\subsection{Initial basis and conventions matching}
%Running the IBPs, here we used \texttt{LiteRed2}~\cite{Lee:2013mka}, following the considerations in~\cite{Gorges:2023zgv,Duhr:2025lbz,Maggio:2025jel}, we pick the following starting basis
\beq\bsp\label{eq:initial_basis}
 I_1=&I_{0, 1, 1, 1, 0, 0, 0, 0, 0}\,, \\
 I_2=&I_{1, 0, 1, 1, 0, 0, 0, 0, 0}\,, \\
 I_3=&I_{1, 1, 1, 1, 0, 0, 0, 0, 0}\,, \\
 I_4=&I_{2, 1, 1, 1, 0, 0, 0, 0, 0}\,, \\
 I_5=&I_{1, 2, 1, 1, 0, 0, 0, 0, 0}\,, \\
 I_6=&I_{1, 1, 1, 1, -1, 0, 0, 0, 0}\,, \\
 I_7=&I_{3, 1, 1, 1, 0, 0, 0, 0, 0}\,.
\esp\eeq
The computation of any member of the family defined by eq.~\eqref{eq:banana_family_def} is then reduced to the computation of the master integrals.
The master integrals $I_1$ and $I_2$ are products of one-loop tadpole integrals, and we will not discuss them any further. The remaining master integrals $I_k$ with $3\le k\le 7$ are genuine three-loop integrals, and their computation is the main focus of this paper. In the following it will be useful to group the master integrals into a vector,
\beq
\bI = \big(I_1,\ldots,I_7\big)^T\,.
\eeq
%Note that there is some arbitrariness in how we choose the basis of master integrals. We will come back to this point below. 

In order to compute the master integrals, we use the method of differential equations~\cite{Kotikov:1990kg,Kotikov:1991hm,Kotikov:1991pm,Gehrmann:1999as}. 
If we compute the derivatives of $\bI$ with respect to $z_i$, we again obtain a linear combination of master integrals, i.e., we obtain linear systems of equations,
\beq\label{eq:initial_DEQ}
\partial_{z_i}\bI(\bz,\eps) = \bOmega_i(\bz,\eps)\bI(\bz,\eps)\,, \qquad i=1,2\,,
\eeq
where $\bz=(z_1,z_2)$ and $\bOmega_i(\bz,\eps)$ are matrices of rational functions in $\bz$ and $\eps$. At this point we note that there is some arbitrariness in our choice of basis in eq.~\eqref{eq:initial_basis}. Indeed, a judicious choice of master integrals may simplify finding a solution of the differential equations in eq.~\eqref{eq:initial_DEQ}. If we define a new basis of master integrals by $\bM(\bz,\eps) = \bR(\bz,\eps)\bI(\bz,\eps)$, then the vector $\bM(\bz,\eps)$ satisfies differential equations where the matrices $\bs{\widetilde{\Omega}}_i(\bz,\eps)$ are related to the $\bOmega_i(\bz,\eps)$ by a gauge transformation,
\beq
 \bs{\widetilde{\Omega}}_i = \big(\partial_{z_i}\bR + \bR\bOmega_i\big)\bR^{-1}\,.
 \eeq
A particularly convenient choice is a so-called \emph{canonical basis}. They were originally introduced in ref.~\cite{Henn:2013pwa} in the context of differential equations for Feynman integrals that evaluate to multiple polylogarithms~\cite{Mpls1}, based on the concepts of uniform transcendental weight~\cite{Kotikov:2010gf} and pure functions~\cite{Arkani-Hamed:2010pyv} from $\mathcal{N}=4$ Super Yang-Mills theory. 
Here we use the definition of a canonical basis beyond multiple polylogarithms of refs.~\cite{Pogel:2022yat,Pogel:2022ken,Pogel:2022vat,Gorges:2023zgv,Duhr:2025lbz,Maggio:2025jel,e-collaboration:2025frv,Bree:2025tug}. In particular, the method of refs.~\cite{Gorges:2023zgv,Duhr:2025lbz} is based on an extension of the concepts of uniform transcendental weights and pure functions beyond polylogarithms introduced in ref.~\cite{Broedel:2018qkq}. A main feature of a canonical basis is that it satisfies an $\eps$-factorised differential equation,
\beq\label{eq:DEQ_can_basis}
\rd \bM(\bs{\uz},\eps) = \eps\,\bA(\bs{\uz})\bM(\bs{\uz},\eps)\,,
\eeq
where $\rd = \rd z_1\,\partial_{z_1} + \rd z_2\,\partial_{z_2}$ is the total differential.
The solution of an $\eps$-factorised differential equation  can be expressed as
\begin{equation}
\bM(\bx,\eps) = \mathbb{P}_{\gamma}(\bs{\uz},\eps)\bM_0(\eps)\,,
\end{equation}
where $\bM_0(\eps)$ denotes the value of $\bM(\bs{\uz},\eps)$ at some point $\bs{\uz}=\bs{\uz}_0$ and $\gamma$ is a path from $\bs{\uz}_0$ to a generic point $\bs{\uz}$. We denote by $\mathbb{P}_{\gamma}(\bs{z},\eps)$ the path-ordered exponential 
\begin{equation}\label{eq:Pexp_def}
\mathbb{P}_{\gamma}(\bs{\uz},\eps) = \mathbb{P}\exp\left[\eps\int_{\gamma}\bA(\bs{\uz})\right]\,,
\end{equation}
which can easily be expanded in $\eps$ up to any given order, and the coefficients of this expansion involve iterated integrals~\cite{ChenSymbol},
\begin{equation}\label{eq:Pexp_def_exp}
\mathbb{P}_{\gamma}(\bs{\uz},\eps) = \mathds{1} + \sum_{k=1}^\infty \eps^k\sum_{1\le i_1,\ldots,i_k\le p}\bA_{i_1}\cdots\bA_{i_k}  I_{\gamma}(\omega_{i_1},\ldots,\omega_{i_k}) \,,
\end{equation}
with
\beq\label{eq:A_basis_omega}
\bA(\bs{\uz}) = \sum_{i=1}^p \bA_i \omega_i\,,
\eeq
and where we defined the iterated integral
\begin{equation}\bsp\label{eq:iterated_int_def}
 I_{\gamma}(\omega_{i_1},\ldots,\omega_{i_k}) &\,= \int_{\gamma} \omega_{i_1}\cdots \omega_{i_k}\\
 &\,= \int_{0\le \xi_k\le \cdots \le \xi_1\le1}\rd \xi_1 \,f_{i_1}(\xi_1)\,\rd \xi_2 \,f_{i_2}(\xi_2)\cdots \rd \xi_k\,f_{i_k}(\xi_k)\,,
\esp\end{equation}
with $\gamma^*\omega_{i_r}=\rd\xi_r\,f_{i_r}(\xi_r)$ the pullback of $\omega_i$ to the path $\gamma$. The $\omega_i$ are called the \emph{letters} of the iterated integrals, and the number $k$ of letters is the \emph{length}.

Canonical bases have been obtained for three-loop banana integrals for various mass assignments in refs.~\cite{Pogel:2022yat,Maggio:2025jel,Duhr:2025kkq,Pogel:2025bca}. While it is by now well established how to derive a canonical basis for banana integrals with three (and even more) loops, the required class of functions, in particular the iterated integrals that arise in the $\eps$-expansion of the path-ordered exponential, is not always equally well understood. This can be traced back to the fact that in general $L$-loop banana integrals are associated with a family of Calabi-Yau (CY) $(L-1)$-folds~\cite{MR3780269,Bloch:2014qca,Primo:2017ipr,Bonisch:2020qmm,Bonisch:2021yfw}. When computed in $D=2$ dimensions, the maximal cuts of the banana integrals, which are solutions to the associated homogeneous differential equations~\cite{Primo:2016ebd,Frellesvig:2017aai,Bosma:2017ens}, compute the periods of these CY varieties (and their derivatives). The latter also enter the transformation to the canonical basis and the differential equation matrix. As a consequence, the letters $\omega_i$ do not only involve rational or algebraic functions of the kinematic variables $\bz$, but also the periods of the CY $(L-1)$-fold. The periods are themselves computed as the solutions to a system of (partial) differential equations, and they typically define new classes of transcendental, multi-valued functions that have not been studied in the literature.

If we focus on the case of three-loop banana integrals, then there are notable exceptions. For example, in the equal-mass case, the associated CY twofold (also known as a K3 surface) is the symmetric square of an elliptic curve~\cite{doran,BognerCY,BognerThesis,Bloch:2014qca}. The three periods of this K3 surface can be written as products of periods of the same family of elliptic curves that describe the two-loop equal-mass sunrise integral~\cite{Laporta:2004rb,Joyce1973SimpleCubicLGF,Primo:2017ipr},
\beq\bsp\label{eq:sunrise_periods}
\Psi_1(t)&\, = \frac{4 }{\pi  \sqrt[4]{(t-9) (t-1)^3}}\,\K\!\left(\Lambda(t)\right)\,,\\
\Psi_2(t)&\, = \frac{4 }{\pi  \sqrt[4]{(t-9) (t-1)^3}}\,\K\!\left(1-\Lambda(t)\right)\,,
\esp\eeq
with
\beq
\Lambda(t) = \frac{t^2-6 t+\sqrt{(t-9) (t-1)^3}-3}{2 \sqrt{(t-9) (t-1)^3}}\,,
\eeq
and we defined the complete elliptic integral of the first kind,
\beq
\K(\lambda) = \int_0^1\frac{\rd t}{\sqrt{(1-t^2)(1-\lambda t^2)}}\,.
\eeq
This fact was used in refs.~\cite{Broedel:2019kmn,Broedel:2021zij,Pogel:2022yat} to express the master integrals of this family in terms iterated integrals of meromorphic modular forms~\cite{ManinModular,Brown:2014pnb,Matthes2022IteratedPrimitives,Broedel:2021zij} for the congruence subgroup $\Gamma_1(6)$ (and integrals thereof), defined by
\beq
\Gamma_1(N) = \left\{\left(\begin{smallmatrix} a& b\\c&d\end{smallmatrix}\right) \in \SL_2(\mathbb{Z}) : a,d \equiv 1\!\!\! \mod N \textrm{~~and~~} c \equiv 0\!\!\!\mod N\right\}\,.
\eeq
We define the elliptic modulus in the usual way as the ratio of the two periods,
\beq
\tau = i\,\frac{\Psi_2(t)}{\Psi_1(t)}\,.
\eeq
The inverse is a Hauptmodul for $\Gamma_0(6)$~\cite{Maier},
\beq\label{eq:G06_Hauptmodul}
t(\tau) =9\,\frac{\eta(6\tau)^8\,\eta(\tau)^4}{\eta(2\tau)^8\,\eta(3\tau)^4}\,,
\eeq
with
\beq
\Gamma_0(N) = \left\{\left(\begin{smallmatrix} a& b\\c&d\end{smallmatrix}\right) \in \SL_2(\mathbb{Z}) :  c \equiv 0\!\!\!\mod N\right\}\,,
\eeq
and $\eta(\tau)$ is the Dedekind eta function,
\beq
\eta(\tau) = q^{1/24}\prod_{n=1}^\infty(1-q^n)\,,\qquad q=e^{2\pi i \tau}\,. 
\eeq
In our conventions the (holomorphic) period takes the form 
\beq\label{eq:e.m._period_modular}
\Pi_0^{\textrm{e.m.}}(\tau) = \frac{1}{12}\,(t(\tau)-9)(t(\tau)-1)\Psi_1(t(\tau))^2\,,
\eeq
where $\Psi_1(t(\tau))$ is an Eisenstein series of weight one for $\Gamma_1(6)$:
\beq\label{eq:psi_def}
\Psi_1(t(\tau)) := \frac{2}{\sqrt{3}}\,\frac{\eta(3\tau)\,\eta(2\tau)^6}{\eta(\tau)^3\,\eta(6\tau)^2}\,.\footnote{In the following, we will drop the $\tau$ dependence in $\Psi_1(t(\tau))$ for notational clarity.}
\eeq
As a consequence, it is possible to express the master integrals for the three-loop equal-mass banana family fully analytically in terms of a class of special functions that has been studied in the mathematics literature.

In the case of unequal (and non-zero) masses, however, the situation is very different: while it is by now clear from refs.~\cite{Pogel:2022yat,Pogel:2022vat,Pogel:2022ken,Gorges:2023zgv,Duhr:2025lbz,Maggio:2025jel,Duhr:2025kkq,Pogel:2025bca} how to find canonical bases for banana integrals, in general it is not known how to express the results in terms of a class of functions that is well studied in the literature. In ref.~\cite{Duhr:2025tdf} it was shown that the periods associated with the families of K3 surfaces associated with three-loop banana integrals with non-zero propagator masses can always be identified with classes of generalised modular forms. Specifically, for the case of three equal masses studied here, the periods are products of periods of two different elliptic curves. This was made explicit in ref.~\cite{Duhr:2025ppd}, where concrete expressions for the periods in terms of complete elliptic integrals and modular forms were presented. More precisely, let us denote the vectors of periods of the family of K3 surfaces by
\beq
\bs{\uPi}(\bs{\uz}) = \big(\psi_0(\bs{\uz}),\psi^{(1)}_1(\bs{\uz}),\psi^{(2)}_1(\bs{\uz}),\psi_2(\bs{\uz})\big)^T\,.
\eeq
It is known that there is a point of maximal unipotent monodromy (MUM) at $\bz=0$~\cite{Klemm:2019dbm,Bonisch:2020qmm}, and we choose the entries of $\bs{\uPi}(\bz)$ such that $\psi_0(\bz)$ is holomorphic at $\bz=0$ and the $\psi^{(1)}_i(\bs{\uz})$ diverge like a single power of a logarithm,
\beq\bsp
    \psi_0(\bs{\uz})=&\,1 + 3\,z_1 +z_2 + 12\,z_1\,z_2+ 15\,(z_1)^2+(z_2)^2 +\ord(z_i^3)\,,\\
    \psi_{1}^{(1)}(\bs{\uz})=&\,\log{(z_1)}\,\psi_0(\bs{\uz})+4\,z_1 +2\,z_2+28\,z_1\,z_2+26\,(z_1)^2+3\,(z_2)^2+\ord(z_i^3)\,,\\
    \psi_{1}^{(2)}(\bs{\uz})=&\,\log{(z_2)}\,\psi_0(\bs{\uz})+6\,z_1 + 45\,(z_1)^2+12\,z_1\,z_2+\ord(z_i^3)\,.
    \label{periodsourconventions}
\esp\eeq
The remaining period is not independent from the first three, and it can be obtained from the quadratic relation
\beq
\bs{\uPi}(\bs{\uz})^T\bs{\Sigma}_{31}\bs{\uPi}(\bs{\uz})=0\,,
\eeq
with the intersection pairing
\beq\label{eq:Sigma_31}
\bs{\Sigma}_{31} = \left(\begin{smallmatrix}
 0 & 0 & 0 & 1 \\
 0 & -6 & -3 & 0 \\
 0 & -3 & 0 & 0 \\
 1 & 0 & 0 & 0 
 \end{smallmatrix}\right).
 \eeq
 We also define the \emph{canonical coordinates} close to $\bs{z}=0$,
 \beq\label{eq:t_def}
\sigma_k(\bs{\uz}) = \frac{\psi_1^{(k)}(\bs{\uz})}{\psi_0(\bs{\uz})} = \log z_k+ \ord(z_l)\,,\quad q_k(\bs{\uz})=e^{\sigma_k(\bs{\uz})}\,.
\eeq
The inverse map, which expresses $\bz$ as a function of the canonical coordinates $\bs{\sigma}=(\sigma_1,\sigma_2)$ is called the \emph{mirror map}. 

In ref.~\cite{Duhr:2025ppd} it was shown that the periods of the K3 surface associated with banana integrals with three equal masses can be expressed as products of the periods of the elliptic curve of the equal-mass sunrise integral in eq.~\eqref{eq:sunrise_periods}. Translated to our conventions, the result reads,
%\beq\bsp\label{eq:K3_to_elliptic}
%    \psi_0(\bs{u})=&-\frac{u_1 \left(u_1^2+2\, u_1-3\right) \left(u_2^3-u_2^2-9\, u_2+9\right) }{12 \left(u_1+1\right) \left(u_1^2-\left(u_2^2-3\right) u_1+u_2^2\right)}\Psi _1\left(u_1\right) \Psi _1\left(u_2\right)\,,\\
%    \psi_{1}^{(1)}(\bs{u})=&\,-\pi\phi_{1}^{(1)}(\bs{u})-i\,\pi\,\psi_0(\bs{u})\,,\\
%    \psi_{1}^{(2)}(\bs{u})=&\,\pi\,\phi_{1}^{(1)}(\bs{u})-2\,\pi\,\phi_{1}^{(2)}(\bs{u})-i\,\pi\,\psi_0(\bs{u})\,,
%\esp\eeq
%where we defined
%\beq\bsp
%    \phi_{1}^{(1)}(\bs{u})=&-\frac{u_1 \left(u_1^2+2\, u_1-3\right) \left(u_2^3-u_2^2-9\, u_2+9\right)  }{12 \left(u_1+1\right) \left(u_1^2-\left(u_2^2-3\right) u_1+u_2^2\right)}\Psi _2\left(u_1\right)\Psi _1\left(u_2\right)\,,\\
%    \phi_{1}^{(2)}(\bs{u})=&-\frac{u_1 \left(u_1^2+2\, u_1-3\right) \left(u_2^3-u_2^2-9\, u_2+9\right)}{12 \left(u_1+1\right) \left(u_1^2-\left(u_2^2-3\right) u_1+u_2^2\right)} \Psi _1\left(u_1\right) \Psi _2\left(u_2\right)\,.
%\esp\eeq
\begin{align}\label{eq:K3_to_elliptic}
\psi_0(\bs{u})=&-\frac{u_1 \left(u_1^2+2\, u_1-3\right) \left(u_2^3-u_2^2-9\, u_2+9\right) }{12 \left(u_1+1\right) \left(u_1^2-\left(u_2^2-3\right) u_1+u_2^2\right)}\Psi _1\left(u_1\right) \Psi _1\left(u_2\right)\,,\\
 \nonumber   \psi_{1}^{(1)}(\bs{u})=&\,\frac{\pi \,u_1 \left(u_1^2+2\, u_1-3\right) \left(u_2^3-u_2^2-9\, u_2+9\right) }{12 \left(u_1+1\right) \left(u_1^2-\left(u_2^2-3\right) u_1+u_2^2\right)}\left[\Psi _2\left(u_1\right)\Psi _1\left(u_2\right)+i \Psi _1\left(u_1\right)\Psi _1\left(u_2\right)\right]\,,\\
\nonumber    \psi_{1}^{(2)}(\bs{u})=&\,\frac{\pi\, u_1 \left(u_1^2+2\, u_1-3\right) \left(u_2^3-u_2^2-9\, u_2+9\right) }{12 \left(u_1+1\right) \left(u_1^2-\left(u_2^2-3\right) u_1+u_2^2\right)}\,\times\\
\nonumber    &\qquad\qquad\times\left[i \Psi _1\left(u_1\right)\Psi _1\left(u_2\right) +2\Psi _1\left(u_1\right)\Psi _2\left(u_2\right)-\Psi _2\left(u_1\right)\Psi _1\left(u_2\right)\right]\,.
\end{align}
The variables $\bs{u}=(u_1,u_2)$ are related to $\bs{\uz}$ by
\beq\bsp\label{eq:z_to_u}
    z_1(\bs{u})&\,=\frac{\left(u_1^2-u_1\, u_2^2+3 \,u_1+u_2^2\right)^2}{(u_1-1)\, u_1\, (u_1+3) (u_2-3) (u_2-1) (u_2+1) (u_2+3)}\,,\\
    z_2(\bs{u})&\,=\frac{(u_1-3)^2 (u_1+1)^2\, u_2^2}{(u_1-1)\, u_1\, (u_1+3) (u_2-3) (u_2-1) (u_2+1) (u_2+3)}\,.
\esp\eeq
In order to simplify the notations, we use the shorthands $\psi_0(\bs{u}) := \psi_0(\bs{\uz}(\bs{u}))$ and $\psi_1^{(k)}(\bs{u}) := \psi_1^{(k)}(\bs{\uz}(\bs{u}))$, and we also defined
$\Psi_i(u_j) := \Psi_i(t_j(u_j))$, with
\beq\label{eq:tj_to_uj}
t_j(u_j) = \frac{u_j(3-u_j)}{1+u_j}\,.
\eeq
Just like for the equal-mass two-loop sunrise and three-loop banana integrals, we then obtain a modular parametrisation for the mirror map and the holomorphic periods. For the mirror map, this is achieved by performing the following identification in eq.~\eqref{eq:z_to_u},
\beq\label{eq:u_modular}
u_1=u\left(\tfrac{\tau_1}{6}\right) \textrm{~~~and~~~} u_2=u\left(\tau_2\right)\,, 
\eeq
where $u(\tau)$ is a Hauptmodul for the congruence subgroup $\Gamma_0(12)$:
\begin{equation}
    u(\tau)=3\frac{\eta(2\tau)^2\eta(12\tau)^4}{\eta(4\tau)^4\eta(6\tau)^2}\,,
\end{equation}
and the elliptic moduli $\bs{\tau} = (\tau_1,\tau_2)$ are related to the canonical coordinates $\bs{\sigma}$ via\footnote{We have swapped the indices of $\tau_1$ and $\tau_2$ with respect to the choice in ref.~\cite{Duhr:2025ppd} for notational convenience.}
\begin{align}\label{eq:tau_to_sigma}
    \tau_1=\sigma_1\textrm{~~~and~~~}
    \tau_2=3\sigma_1+3\sigma_2\,;\quad \tilde{q}_1=e^{\tau_1/6}\textrm{~~~and~~~}\tilde{q}_2=e^{\tau_2}\,.
\end{align}
Note that $u(\tau)$ is related to the Hauptmodul $t(\tau)$ for $\Gamma_0(6)$ in eq.~\eqref{eq:G06_Hauptmodul} via the relation (cf. eq.~\eqref{eq:tj_to_uj})~\cite{Maier},
\begin{equation}
\label{eq:t_to_u}
    t(\tau)=\frac{u(\tau)(3-u(\tau))}{1+u(\tau)}.
\end{equation}
Likewise, the holomorphic period can be identified as a product of Eisenstein series upon inserting eqs.~\eqref{eq:e.m._period_modular} and~\eqref{eq:u_modular} into eq.~\eqref{eq:K3_to_elliptic}. We thus see that the periods and the mirror map for the family of K3 surfaces associated with the three-loop banana integral with three equal masses admit a parametrisation in terms of modular forms and functions for the congruence subgroup $\Gamma_1(12)$. In the remainder of this paper we combine these results from refs.~\cite{Duhr:2025tdf,Duhr:2025ppd} with the canonical form of the differential equation, to obtain an analytic result for all master integrals in terms of iterated integrals over modular forms, similar to the equal-mass case.

\section{Analytic results for three-loop banana integrals with three equal masses}
\label{sec:canonicaldeq}

In the previous section we have reviewed modular parametrisations of the corresponding periods and mirror map. In this section we show how we can combine this result with the canonical differential equations to obtain full analytic expressions for all master integrals in terms of iterated integrals of modular forms. 

\subsection{The canonical differential equations}
We start by defining the rotation from the original basis of master integrals $\bI$ defined in eq.~\eqref{eq:initial_basis} to the canonical basis $\bM$. We consider the master integrals as functions of the two independent variables $\bs{\tau}$ defined in eq.~\eqref{eq:tau_to_sigma}. When working with the method of refs.~\cite{Gorges:2023zgv,Maggio:2025jel,Duhr:2025lbz}, it is paramount to have a good starting basis of master integrals which reflects as much as possible the underlying geometry. Our starting basis in eq.~\eqref{eq:initial_basis} has this property. In particular, the master integrals $I_3$, $I_4$, $I_5$ and $I_7$ at $\eps=0$ compute the periods and quasi-periods of the K3 surface and $I_6$ has a non-vanishing residue at infinity. For more details, we refer to refs.~\cite{Duhr:2025lbz,Maggio:2025jel}. 

After applying the method of refs.~\cite{Gorges:2023zgv,Maggio:2025jel,Duhr:2025lbz}, we find that our original basis from eq.~\eqref{eq:initial_basis} and the canonical basis are related by,
\begin{align}
    &M_1=\eps^3\,I_{011100000}\,,\nonumber\\
    &M_2=\eps^3\,I_{101100000}\,,\nonumber\\
    &M_3=\eps^3\,\frac{I_{111100000}}{\psi_0}\,,\nonumber\\
  \label{eq:canon_rot}  &M_4=\frac{1}{\eps}{\partial_{\tau_2} M_3}-F_{11}\,M_3\,,\\
    &M_5=\frac{1}{\eps}{\partial_{\tau_1} M_3}-F_{12}\,M_3\,,\nonumber\\
    &M_6=\eps^3\,I_{1111-10000}-F_{31}\,M_3\,,\nonumber\\
    &M_7=\frac{1}{\eps}{\partial_{\tau_2} M_5}-F_{21}\,M_3-F_{22}\,M_4-F_{25}\,M_5-F_{34}\,M_6-F_{01}\,M_1-F_{02}\,M_2\,.\nonumber
\end{align}
The functions $F_{ij}(\bs{\tau})$ are defined such that the terms in the differential equations that are not $\eps$-factorised vanish.
This constraint can be expressed as a set of differential equations satisfied by these functions (for details, see refs.~\cite{Gorges:2023zgv,Maggio:2025jel,Duhr:2025lbz}).  Using the modular parametrisation for the periods introduced in the previous section, we can solve these differential equations and we obtain closed expressions for all of them in terms of modular forms and functions, as well integrals over them. We find
\beq\bsp
    F_{11}(\bs{\tau})&=\bbI_1(\bs{\tau})+\frac{1}{2} (\phi_{0,2}^b(\bs{\tau})+ \phi_{0,2}^c(\bs{\tau}))\,,\\
    F_{12}(\bs{\tau})&=\bbI_2(\bs{\tau})+\frac{1}{2} (\phi_{2,0}^b(\bs{\tau})+ \phi_{2,0}^c(\bs{\tau}))\,,\\
    F_{25}(\bs{\tau})&=-\bbI_1(\bs{\tau})+\frac{1}{2} (\phi_{0,2}^b(\bs{\tau})+ \phi_{0,2}^c(\bs{\tau}))\,,\\
    F_{22}(\bs{\tau})&=-2\,\bbI_2(\bs{\tau})\,,\\
    F_{31}(\bs{\tau})&=\bbI_3(\bs{\tau})+ \phi_{1,1}(\bs{\tau})\,,\\%=\partial_{\tau_1}\textrm{II}_4 + f_6
    F_{34}(\bs{\tau})&=-18\,\bbI_3(\bs{\tau})\,,\\
    F_{01}(\bs{\tau})&=21\,\bbI_3(\bs{\tau})\,,\\
    F_{02}(\bs{\tau})&=3\,\bbI_3(\bs{\tau})\,,\\
    F_{21}(\bs{\tau})&=-2 \,\bbI_1 (\bs{\tau})\,\bbI_2(\bs{\tau}) - 
 9 \,\bbI_3(\bs{\tau})^2 + \phi_{2,2}(\bs{\tau})\,.
    %\int\mathrm{d}\tau_1 f_8 + 
 %\partial_{\tau_2}\int\mathrm{d}\tau_1\left(
   %\phi_{0,2}^a^2 - 2 f_1 \phi_{2,0}^a - 18 f_5 f_6 - 
    %(\partial_{\tau_2}\textrm{II}_1)^2 - 
    %2 f_1 \partial_{\tau_1}\textrm{II}_2 - 
    %18 f_5 \partial_{\tau_2}\textrm{II}_3\right)\,,
\esp\eeq
The functions $\phi_{i,j}(\bs{\tau})$ and $\phi_{i,j}^r(\bs{\tau})$ are given in appendix~\ref{app:diff_forms_can_eq}. They are rational functions in $\bs{u}$ and contain powers of $\Psi_1(\tau_k)$. Note that they are functions of $\bs{\tau}$ through eqs.~\eqref{eq:e.m._period_modular} and~\eqref{eq:u_modular}, and so they define modular forms and functions in two variables related to the congruence subgroup $\Gamma^1(12)\times \Gamma_1(12)$, with
\beq
\Gamma^1(N) = \big\{\gamma\in \SL_2(\mathbb{Z}): \gamma^T\in\Gamma_1(N)\big\}\,.
\eeq
The functions $\bbI_k$ are integrals of modular forms, and they are given by
\beq\bsp 
\bbI_1(\bs{\tau})&=\int_{i\infty}^{\tau_2}\mathrm{d\tau_2'}\,\partial_{\tau_1}\phi_{-2,4}(\tau_1,\tau'_2)\,,\\
    \bbI_2(\bs{\tau})&=\int_{i\infty}^{\tau_1}\mathrm{d\tau_1'}\,\partial_{\tau_2}\phi_{4,-2}(\tau'_1,\tau_2)\,,\\ \label{new_int}
    \bbI_3(\bs{\tau})&=\int_{i\infty}^{\tau_2}\mathrm{d}\tau_2'\,\partial_{\tau_1}\phi_{-1,3}(\tau_1,\tau'_2)=\int_{i\infty}^{\tau_1}\mathrm{d\tau_1'}\,\partial_{\tau_2}\phi_{3,-1}(\tau'_1,\tau_2)\,.
    %\textrm{II}_4&=\int\mathrm{d}\tau_2f_7\,,\nonumber\\
\esp\eeq
We see that there are only three functions that are not expressible in terms of rational functions and (derivatives of) periods. These functions were called \emph{$\eps$-functions} in refs.~\cite{Duhr:2025kkq,Duhr:2025xyy}. In refs.~\cite{Duhr:2024uid,Duhr:2025xyy} a method based on intersection theory~\cite{Mastrolia:2018uzb,yoshida_hypergeometric_1997,aomoto_theory_2011,Mizera:2017rqa,Mizera:2019gea,matsumoto_relative_2019-1,Frellesvig:2017aai,Frellesvig:2020qot,Frellesvig:2019kgj,Weinzierl:2020xyy,Weinzierl:2020nhw,Cacciatori:2021nli,Chestnov:2022xsy,Brunello:2023rpq,Frellesvig:2019uqt,Fontana:2023amt,Brunello:2024tqf,Crisanti:2024onv,Lu:2024dsb,
Caron-Huot:2021xqj,Caron-Huot:2021iev} was introduced to identify relations between $\eps$-functions (see also ref.~\cite{Pogel:2024sdi} for a related method). In particular, this method has allowed us to identify $F_{21}(\bs{\tau})$ as a quadratic polynomial in the $\bbI_k$ (it can also be seen as a special case of a relation for the banana integral with four unequal masses discussed in ref.~\cite{Duhr:2025kkq}). We note that, up to an overall factor, the expansion of the $\bbI_i$ has integer coefficients when written in the canonical variables $q_1,q_2$ close to the MUM-point. This is similar to the observations made in the equal-mass case~\cite{Pogel:2022yat} (or more general one-parameter families of K3 surfaces~\cite{Duhr:2025lbz}), where the $\eps$-functions are given as primitives of magnetic modular forms~\cite{magnetic1,magnetic3,Bonisch:2024nru,magnetic4}.

Let us discuss some further properties of our $\eps$-functions. We focus on $\mathbb{I}_3(\bs{\tau})$, but we can easily extend the discussion to the other two $\eps$-functions. First, from the explicit expressions in appendix~\ref{app:diff_forms_can_eq}, we see that $\phi_{-1,3}$ is a meromorphic modular form in two variables for $\Gamma^1(12)\times \Gamma_1(12)$ with a simple pole on the locus
\beq\label{eq:pole_locus_1}
u_1^2 + u_2^2 - u_1 ( u_2^2-3) = 0\,.
\eeq 
The derivative $\partial_{\tau_1}\phi_{-1,3}$ then has a double pole on this locus. However, since we have taken the derivative in $\partial_{\tau_1}$, we have lost modularity in $\tau_1$. Hence, $\partial_{\tau_1}\phi_{-1,3}$ is a meromorphic modular form of weight three in $\tau_2$ for $\Gamma_1(12)$ with a double pole on the locus in eq.~\eqref{eq:pole_locus_1}. Since $\mathbb{I}_3$ is obtained by taking a primitive in $\tau_2$, we conclude that $\mathbb{I}_3$ has a simple pole on the locus in eq.~\eqref{eq:pole_locus_1}. Using the same argument for the other modular forms, we see that all our $\eps$-functions only have simple poles.
Second, we find that the following relation among $\eps$-functions holds:
\begin{align}
&\partial_{\tau_1}\bbI_1(\bs{\tau})=\partial_{\tau_2}\bbI_2(\bs{\tau})\,.
%&\textrm{II}_3(\bs{\tau})=\partial_{\tau_1}\int_{i\infty}^{\tau_2}\mathrm{d\tau_2'}\,f_7(\tau_1,\tau_2')\,.
    %&F_{21}=\int\mathrm{d}\tau_2 f_9 + 
 %\partial_{\tau_1}\int\mathrm{d}\tau_2\left(-2 \phi_{0,2}^a f_3 + \phi_{2,0}^a^2 - 18 f_6 f_7 - 
    %2 f_3 \partial_{\tau_2}\textrm{II}_1 - 
    %(\partial_{\tau_1}\textrm{II}_2)^2 - 
    %18 f_7 \partial_{\tau_1}\textrm{II}_4\right).
\end{align}
This relation allows us to obtain alternative representations for $\mathbb{I}_1$ and $\mathbb{I}_2$  as double integrals in the other variable,
\beq\bsp\label{new_int_2}
\mathbb{I}_1(\bs{\tau})&\,=\int_{i\infty}^{\tau_1}\mathrm{d\tau_1'}\int_{i\infty}^{\tau_1'}\mathrm{d\tau_1''}\,\partial_{\tau_2}^2\phi_{4,-2}(\tau''_1,\tau_2)\,,\\
\mathbb{I}_2(\bs{\tau})&\,=\int_{i\infty}^{\tau_2}\mathrm{d\tau_2'}\int_{i\infty}^{\tau_2'}\mathrm{d\tau_2''}\,\partial_{\tau_1}^2\phi_{-2,4}(\tau_1,\tau''_2)\,.
\esp\eeq
Said differently, we see that we can interpret $\mathbb{I}_1$ either via eq.~\eqref{new_int} as the first primitive  in $\tau_2$ of  the modular form $\partial_{\tau_1}\phi_{-2,4}(\tau_1,\cdot)$ with a double pole, or via eq.~\eqref{new_int_2} as the second primitive  in $\tau_1$ of  the modular form $\partial_{\tau_2}^2\phi_{4,-2}(\cdot,\tau_2)$ with a triple pole.

Let us now discuss the structure of the canonical differential equations.
After rotation to the canonical basis, the differential equations take the form
\begin{equation}
\label{eq:can_tau}
\mathrm{d}\bs{M}(\bs{\tau},\eps) = \eps\,\bA(\bs{\tau})\,\bs{M}(\bs{\tau},\eps)\,, 
\end{equation}
with
\begin{equation}
\label{eq:A_tau}
\bA(\bs{\tau})=\bA_1(\bs{\tau})\,\mathrm{d}\tau_1+\bA_2(\bs{\tau})\,\mathrm{d}\tau_2\,,
\end{equation}
and
\begin{equation}\bsp
\label{mat:canonicalforms}
\bA_{i}=\left(
    \begin{array}{ccccccccccccccc}
      \omega^i_{11}&0&0 &0&0&0&0\\
      0&\omega^i_{22}&0 &0&0&0&0\\
      0&0&\omega^i_{33} &\delta_{i2}&\delta_{i1}&0&0\\
       \omega^i_{41}&\omega^i_{42}&\omega^i_{43}&\omega^1_{44}&\omega^i_{45}&\omega^i_{46}&\delta_{i1}\\
       \omega^i_{51}&\omega^i_{52}&\omega^i_{53} &\omega^1_{54}&\omega^i_{55}&\omega^i_{56}&\delta_{i2}\\
\omega^i_{61}&\omega^i_{62} &\omega^i_{63}&\omega^i_{64}&\omega^i_{65}&\omega^i_{66}&0\\
\omega^i_{71}&\omega^i_{72} &\omega^i_{73}&\omega^i_{74}&\omega^i_{75}&\omega^i_{76}&\omega^i_{77}
    \end{array}\right)\,,
\esp\end{equation}
% \begin{equation}\bsp
% \label{mat:canonicalforms}
% \bA_{1}=\left(
%     \begin{array}{ccccccccccccccc}
%       \omega^1_{1,1}&0&0 &0&0&0&0\\
%       0&\omega^1_{2,2}&0 &0&0&0&0\\
%       0&0&\omega^1_{3,3} &1&0&0&0\\
%        \omega^1_{4,1}&\omega^1_{4,2}&\omega^1_{4,3}&\omega^1_{4,4}&\omega^1_{4,5}&\omega^1_{4,6}&0\\
%        \omega^1_{5,1}&\omega^1_{5,2}&\omega^1_{5,3} &\omega^1_{5,4}&\omega^1_{5,5}&\omega^1_{5,6}&1\\
% \omega^1_{6,1}&\omega^1_{6,2} &\omega^1_{6,3}&\omega^1_{6,4}&\omega^1_{6,5}&\omega^1_{6,6}&0\\
% \omega^1_{7,1}&\omega^1_{7,2} &\omega^1_{7,3}&\omega^1_{7,4}&\omega^1_{7,5}&\omega^1_{7,6}&\omega^1_{7,7}
%     \end{array}\right)\,,\\
% \bA_{2}=\left(
%     \begin{array}{ccccccccccccccc}
%       \omega^2_{1,1}&0&0 &0&0&0&0\\
%       0&\omega^2_{2,2}&0 &0&0&0&0\\
%       0&0&\omega^2_{3,3} &0&1&0&0\\
%        \omega^2_{4,1}&\omega^2_{4,2}&\omega^2_{4,3}&\omega^2_{4,4}&\omega^2_{4,5}&\omega^2_{4,6}&1\\
%        \omega^2_{5,1}&\omega^2_{5,2}&\omega^2_{5,3} &\omega^2_{5,4}&\omega^2_{5,5}&\omega^2_{5,6}&0\\
% \omega^2_{6,1}&\omega^2_{6,2} &\omega^2_{6,3}&\omega^2_{6,4}&\omega^2_{6,5}&\omega^2_{6,6}&0\\
% \omega^2_{7,1}&\omega^2_{7,2} &\omega^2_{7,3}&\omega^2_{7,4}&\omega^2_{7,5}&\omega^2_{7,6}&\omega^2_{7,7}
%     \end{array}\right)\,,
% \esp\end{equation}
 where $\delta_{ij}$ is the Kronecker delta and the $\omega^k_{ij}$ are rational functions in $u_i$, the holomorphic periods $\Psi_1(\tau_k)$ and the $\eps$-functions defined in eq.~\eqref{new_int}. Their explicit expressions are provided in appendix~\ref{app:diff_forms_can_eq}. 

%%%%%%%%%%%%%%%%%%%%%%%%%%%%%%%%%%
\subsection{Properties of the canonical differential equations}
In the previous subsection we have presented the canonical differential equations for the three-loop banana integrals with three equal masses, and we have discussed the associated $\eps$-functions. The rotation to the canonical basis in eq.~\eqref{eq:canon_rot} was constructed using the method of refs.~\cite{Gorges:2023zgv,Maggio:2025jel,Duhr:2025lbz}, which aims at constructing a rotation that achieves $\eps$-factorisation. In ref.~\cite{Duhr:2025lbz} it was proposed that the resulting differential equations enjoy additional properties that go beyond mere $\eps$-factorisation. In particular, it was argued in ref.~\cite{Duhr:2025lbz} that the differential forms $\omega_{ij}$ should have at most simple poles and they should define independent cohomology classes. While these properties are manifest for canonical systems involving dlog-forms, checking them for cases with non-trivial geometries is a complicated task. 
In the following we show how we can leverage the fact that the differential forms $\omega^k_{ij}$ are constructed from meromorphic modular forms in two variables to check that the properties conjectured in ref.~\cite{Duhr:2025lbz} hold for the three-loop banana integrals with three equal masses.

\paragraph{Simple poles.} From the explicit expressions for the functions $\omega_{ij}^k$ in appendix~\ref{app:diff_forms_can_eq}, we can see that the $\omega^k_{ij}$ have poles at $\tilde{q}_k=0$ and/or on the codimension-one loci\footnote{There are also apparent singularities at $u_k=-1$. However, these singularities are spurious and cancel against a zero in $\Psi_1(u_k)$; see the discussion in appendix~\ref{app:modular}.}
$L_1(\bs{u})=0$ or $L_2(\bs{u})=0$, where we defined
\beq\bsp
L_1(\bs{u})&\, = u_1^2 + u_2^2 - u_1 ( u_2^2-3)\,,\\
L_2(\bs{u})&\,=(u_2 u_1-3 u_1-3 u_2-3) (u_2 u_1+3 u_1-3 u_2+3)\\&\,\times (u_2 u_1-u_1+u_2+3) (u_2 u_1+u_1+u_2-3)\,.
\esp\eeq
Using the explicit expressions in appendix~\ref{app:diff_forms_can_eq}, we have checked that the differential forms $\rd\tau_k\,\omega_{ij}^k$ only have simple poles in $\tau_k$. We stress that this feature arises in a very non-trivial manner, because some of the meromorphic modular forms in appendix~\ref{app:diff_forms_can_eq} have double or even triple poles. These apparent higher-order poles cancel in the specific combinations that define the $\omega_{ij}^k$. As an example, consider the function 
\beq\label{eq:omega_75_1_example}
\omega_{75}^1 = -2 \bbI_1
    \bbI_2 - 9 
\bbI_3^2 + \phi_{2,2}\,.
\eeq
From the explicit expressions in appendix~\ref{app:diff_forms_can_eq}, we see that $\phi_{2,2}$ has double poles on the loci $L_1(\bs{u})=0$ and $L_2(\bs{u})=0$. All three $\eps$-functions $\mathbb{I}_k$ $k\in\{1,2,3\}$, have simple poles on the locus $L_1(\bs{u})=0$. $\mathbb{I}_1$ and $\mathbb{I}_2$ also have a simple on the locus $L_2(\bs{u})=0$, while $\mathbb{I}_3$ is regular there. One can easily check that the combination in eq.~\eqref{eq:omega_75_1_example} only has simple poles on the loci $L_1(\bs{u})=0$ and $L_2(\bs{u})=0$.

So far we have only discussed the pole structure for codimension-one loci. Since we deal with differential forms in two variables, we also need to investigate the behaviour at poles on codimension-two loci. For example, take a function $\omega_{ij}^1$ that has a simple pole in $\tau_1$ on the codimension locus $L_1(\bs{u})=0$. Explicitly, the poles in the variable $\tau_1$ are given by those values of $u_1$ such that,
\beq
u_1 = u_1^{\pm}(u_2) := \frac{1}{2} \left(u_2^2-3\pm\sqrt{(u_2^2-9) \
(u_2^2-1)}\right)\,.
\eeq
As $u_2$ varies, we obtain two branches of the codimension-one locus $L_1(\bs{u})=0$, given by $u_1=u_1^{+}(u_2)$ and $u_1=u_1^{-}(u_2)$. However, on the codimension-two locus inside $L_1(\bs{u})=0$ defined by values of $u_2$ such that $u_1^{+}(u_2)=u_1^{-}(u_2)$, the two zeroes of $L_1$ coincide, and we obtain a double-pole in $\tau_1$. This happens precisely for $u_2\in\{\pm1,\pm3\}$. Hence, one the codimension-two locus $\bs{u}\in\{(-1,\pm1),(3,\pm3)\}$, the differential forms $\rd\tau_1\,\omega_{ij}^1$ have double poles in $\tau_1$, apparently contradicting the expectation that our differential form should only have simple poles. We find that $\omega_{ij}^1$ always has a numerator that vanishes on the codimension-two locus $\bs{u}\in\{(-1,\pm1),(3,\pm3)\}$, thereby reducing the order of the pole.

\paragraph{Independence of cohomology classes.} We have also checked that the differential forms $\omega_{ij}$ define independent cohomology classes. Loosely speaking, this means that, if we pick a linearly independent set of differential forms $\omega^k_{ij}$, then there is no linear-combination that equals a total derivative (for details, see appendices~\ref{app:modular} and~\ref{app:independence}). The detailed proof is rather lengthy and technical, and it is presented in appendix~\ref{app:independence}. Here it suffices to say that our proof draws heavily on the properties of meromorphic modular forms~\cite{Matthes2022IteratedPrimitives,Broedel:2021zij} and the properties of $\eps$-functions, in particular the fact that we can represent them in different ways as primitives in either $\tau_1$ or $\tau_2$ (cf.~eqs.~\eqref{new_int} and~\eqref{new_int_2}).

\paragraph{Modular properties of the differential forms.} Since our differential forms $\omega_{ij}$ are constructed from meromorphic modular forms in two variables for $\Gamma^1(12)\times\Gamma_1(12)$, it is natural to ask what are their properties under modular transformations. This is motivated by the observations of refs.~\cite{Adams:2018yfj,Bogner:2019lfa,Weinzierl:2020fyx,Pogel:2022yat}, where it was observed in the context of the two-loop sunrise integrals and the three-loop equal-mass banana integrals that the entries of the corresponding canonical differential matrices have a well-defined behaviour under modular transformations. Here we extend these observations to the three-loop banana integral with three equal-masses. This is also the first time that modular properties of Feynman integrals involving modular forms in two variables are studied.

The group $\Gamma^1(12)\times\Gamma_1(12)$ acts on $\bs{\tau}$ via M\"obius transformations,
\beq
(\tau_1,\tau_2) \mapsto (\gamma_1\cdot\tau_1,\gamma_2\cdot\tau_2)\,,
\eeq
with $\gamma_i = \left(\begin{smallmatrix}a_i&b_i\\c_i&d_i\end{smallmatrix}\right)$ and $\gamma_i\cdot\tau_i = \tfrac{a_i\tau_i+b_i}{c_i\tau_i+d_i}$.
A meromorphic function $f:\mathbb{H}^2\to\mathbb{C}$ (with $\mathbb{H}$ the compelx upper half-plane) is a meromorphic modular form in two variables of weights $(k_1,k_2)$ for $\Gamma^1(12)\times\Gamma_1(12)$ if for all $(\gamma_1,\gamma_2)\in \Gamma^1(12)\times\Gamma_1(12)$ we have
\beq\label{eq:modular_2}
f(\gamma_1\cdot\tau_1,\gamma_2\cdot\tau_2) = (c_1\tau_1+d_1)^{k_1}\,(c_2\tau_2+d_2)^{k_2}\,f(\tau_1,\tau_2)\,.
\eeq
The functions $\phi_{l,m}$ and $\phi_{l,m}^r$ defined in appendix~\ref{app:diff_forms_can_eq} define meromorphic modular forms of weights $(l,m)$ for $\Gamma^1(12)\times\Gamma_1(12)$. The $\omega_{ij}^k$ are polynomials in the $\phi_{l,m}$ and $\phi_{l,m}^r$ and our $\eps$-functions. The latter have a more complicated transformation behaviour under modular transformations.\footnote{This transformation behaviour was called \emph{quasi-Eichler} in ref.~\cite{Pogel:2022yat}.} Nevertheless, we find that, if we define the weights of $\mathbb{I}_1$, $\mathbb{I}_2$ and $\mathbb{I}_3$ to be $(0,2)$, $(2,0)$ and $(1,1)$ respectively, then each $\omega_{ij}^k$ has a well-defined weight. We summarise this finding in the following matrices,
\begin{equation}\bsp
\label{mat:weights}
W_{\bA_{1}}=\left(
    \begin{array}{ccccccccccccccc}
      2&0&0 &0&0&0&0\\
      0&2&0 &0&0&0&0\\
      0&0&2 &0&0&0&0\\
       1&1&2 &2&0&1&0\\
       3&3&4&4&2&3&0\\
2&2 &3&3&1&2&0\\
3&3 &4&4&2&3&2
    \end{array}\right)\,,\qquad
W_{\bA_{2}}=\left(
    \begin{array}{ccccccccccccccc}
      2&0&0 &0&0&0&0\\
      0&2&0 &0&0&0&0\\
      0&0&2 &0&0&0&0\\
       3&3&4&2&4&3&0\\
       1&1&2 &0&2&1&0\\
2&2 &3&1&3&2&0\\
3&3 &4&2&4&3&2
    \end{array}\right)\,,
\esp\end{equation}
where we collect the weight of $\omega_{ij}^k$ in $\tau_k$ in the entry $(i,j)$ of $W_{\bA_{k}}$. Note that $W_{\bA_{1}}$ and $ {W_{\bA_{2}}}$ are identical up to exchanging the rows and columns 4 and 5. This is expected, because $M_4$ and $M_5$ are respectively derivatives of $M_3$ with respect to $\tau_2$ and $\tau_1$.
We can also zoom in on the matrices and consider the rows and the columns corresponding to $M_3,M_4,M_5,M_7$, which for $\eps=0$ are related to the periods and quasi-periods of the underlying K3 surface. We have
\begin{equation}\bsp
\label{mat:weights_limit}
W_{\bA_{1}}^{\textrm{K3}}=\left(
    \begin{array}{ccccccccccccccc}
      2 &0&0&0\\
       2 &2&0&0\\
       4&4&2&0\\
        4&4&2&2
    \end{array}\right)\,,\qquad
W_{\bA_{2}}^{\textrm{K3}}=\left(
    \begin{array}{ccccccccccccccc}
      2 &0&0&0\\
       4&2&4&0\\
       2 &0&2&0\\
        4&2&4&2
    \end{array}\right)\,.
\esp\end{equation}
We observe a similar increase in weight from the top right to the bottom left as in other integrals with underlying geometry related to modular forms~\cite{Adams:2018yfj,Bogner:2019lfa,Weinzierl:2020fyx,Pogel:2022yat}.
In particular, we note that by taking the equal-mass limit we recover the same weight structure as for the equal-mass banana considered in ref.~\cite{Pogel:2022yat}. 

So far we have only considered the weights of $\omega_{ij}^k$ for fixed values of $k$. What enters the canonical differential equation in eq.~\eqref{eq:can_tau}, however, are the differential one-forms $\omega_{ij} = \rd\tau_1\,\omega_{ij}^1+\rd\tau_2\,\omega_{ij}^2$, and we conclude by discussing their modular weights. We start by noting that $\rd\tau_1$ and $\rd\tau_2$ transform like modular forms of weights $(-2,0)$ and $(0,-2)$, respectively. By inspecting the explicit expressions in appendix~\ref{app:diff_forms_can_eq}, we see that $\omega_{ij}$ transform  with well-defined weights $(k_{1},k_{2})$ for $\Gamma^1(12)\times\Gamma_1(12)$.
This is equivalent to the statement that the functions $\omega_{ij}^1$ and $\omega_{ij}^2$ transform with weights $(k_1+2,k_2)$ and $(k_1,k_2+2)$, respectively. The weights of the different non-zero entries of $\bA$ are summarised in the following matrix (we indicate by -- a zero entry):
\begin{equation}
\left(
\begin{array}{ccccccc}
 (0,0) & - & - & - & - & - & - \\
 - & (0,0) & - & - & - & - & - \\
 - & - & (0,0) & (-,0) & (0,-) & - & - \\
 (1,3) & (1,3) & (2,4) & (0,0) & (0,4) & (1,3) & (0,-) \\
 (3,1) & (3,1) & (4,2) & (4,0) & (0,0) & (3,1) & (-,0) \\
 (2,2) & (2,2) & (3,3) & (3,1) & (1,3) & (0,0) & - \\
 (3,3) & (3,3) & (4,4) & (4,2) & (2,4) & (3,3) & (0,0) \\
\end{array}
\right)\,.
\end{equation}

\paragraph{Closedness of the differential forms.}

As a consequence of the $\eps$-factorisation, the matrix $\bA$ in eq.~\eqref{eq:A_tau} satisfies the flatness conditions
\beq\label{eq:flatness}
\rd\bA(\bs{\tau}) =  \bA(\bs{\tau}) \wedge  \bA(\bs{\tau}) = 0\,.
\eeq
It follows that the differential forms $\omega_{ij}$ are closed, $\rd\omega_{ij}=0$. This is equivalent to the statement
\beq\label{eq:closed}
\partial_{\tau_2}\omega_{ij}^1 = \partial_{\tau_1}\omega_{ij}^2\,.
\eeq
This equation implies that $\omega_{ij}^1$ and $\omega_{ij}^2$ are not independent. We now argue that eq.~\eqref{eq:closed} implies that at least one of the two functions $\omega_{ij}^1$ or $\omega_{ij}^2$ must contain an $\eps$-function, unless $\omega_{ij}$ has modular weights $(0,0)$.

To see that this is true, let us assume that $\omega_{ij}$ has weights $(k_1,k_2)$ and that neither $\omega_{ij}^1$ or $\omega_{ij}^2$ contain an $\eps$-function. This implies that $\omega_{ij}^1$ and $\omega_{ij}^2$ are modular forms in two variables for $\Gamma^1(12)\times\Gamma_1(12)$ with weights $(k_1+2,k_2)$ and $(k_1,k_2+2)$, respectively. The derivative of a modular form of weight $k\neq0$ is not modular (instead it is quasi-modular; see appendix~\ref{app:modular}). Hence, unless the weight is $(k_1,k_2)=(0,0)$, the left-hand side of eq.~\eqref{eq:closed} would be modular in $\tau_1$ but not in $\tau_2$, while the right-hand side would be modular in $\tau_2$ but not in $\tau_1$, which is a contradiction. It is easy to check that all the $\omega_{ij}$ in appendix~\ref{app:diff_forms_can_eq} of weights $(k_1,k_2)\neq(0,0)$ involve an $\eps$-function.

%The equations~\eqref{eq:can_tau},~\eqref{eq:A_tau} and~\eqref{mat:canonicalforms} (together with the explicit expressions in appendix~\ref{app:diff_forms_can_eq}) are our final results for the canonical differential equations. In the remainder of this subsection, we use the fact that the differential forms $\omega_{ij}^k$ that define the connection matrix $\bA(\bs{\tau})$ can be expressed in terms of modular forms (and integrals thereof)  to study properties of this system.

\subsection{Analytic results for the master integrals}

We can easily write down the general solution to the differential equation in eq.~\eqref{eq:can_tau} in terms of the path-ordered exponential in eq.~\eqref{eq:Pexp_def_exp}. In order to fully specify the solution that we are interested in, we also need to determine the initial condition and the path along which we want to integrate. Since the master integrals $M_1$ and $M_2$ are just products of one-loop tadpole integrals, we do not discuss them any further, and we focus on the master integrals $M_k$ with $3\le k\le 7$. 

A natural choice of boundary condition is the small-mass limit, $m_1^2=m_2^2=0$, which corresponds to $\bs{\sigma}=0$, or equivalently $\bs{\tau}=0$. We stress that there are logarithmic divergences as the masses tend to zero, and so we compute asymptotic expansions around the point $m_1^2=m_2^2=0$. For the master integral $M_3$, this is easily done using the  Mellin-Barnes representation for the integral (cf.,~e.g.,~refs.~\cite{Broedel:2019kmn,Bonisch:2021yfw,Pogel:2022yat}). We find for the first two orders,  
\begin{align}
\label{eq:M3bc}
    M_3=&\eps^3 \Big[16 \zeta_3-\log^2{q_1} (\log{q_1} + 3 \log{q_2}) 
    \Big]\nonumber \\
    &+ 
 \frac{3}{2}\eps^4 \Big[\log^4{q_1} + 2 \log^3{q_1} \log{q_2} + 
   \log^2{q_1} \log^2{q_2} + 
   4 \log{q_1} (\log{q_1} + \log{q_2}) \zeta_2 \nonumber \\
    &+ 
   12 (3 \log{q_1} + \log{q_2}) \zeta_3 + 
   16 \zeta_4\Big]+\mathcal{O}(\eps^5)+\mathcal{O}(q_i)\,,
\end{align}
where we recall that $q_i = e^{\sigma_i}$. From eq.~\eqref{eq:M3bc} we can get the asymptotic behaviour for $M_4,\,M_5$ and $M_7$, because those master integrals are proportional to derivatives of $M_3$ (and to some linear combination of master integrals already determined).
Finally, to obtain the asymptotic expansion for $M_6=\eps^3\,I_{6}-F_{31}\,M_3\,$, we need to compute the expansion of $I_{6}$. This is again done using the Mellin-Barnes representation, and we find
\begin{align}
    M_6= &\frac{4}{3} - \eps (3 \log{q_1} + 
    \log{q_2}) + \eps^2 \Bigg(\frac{7}{2} \log^2{q_1} + 
    2 \log{q_1} \log{q_2} + \frac{1}{2} \log^2{q_2} + 
    2 \zeta_2\Bigg) \nonumber\\
    &- \frac{\eps^3}{6} \Bigg[ 13 \log^3{q_1} + 12 \log^2{q_1} \log{q_2} + 
   6 \log{q_1} \log^2{q_2} + \log^3{q_2} \nonumber\\
    &+ 
   9 (3 \log{q_1} + \log{q_2}) \zeta_2 + 
   24 \zeta_3\Bigg]+\mathcal{O}(\eps^4)+\mathcal{O}(q_i)\,.
\end{align}
In this way we have obtained the leading behaviour of all master integrals in the small-mass limit.

Next let us discuss our choice of path of integration. As a consequence of the flatness condition in eq.~\eqref{eq:flatness}, the result of the path ordered-integral does not depend on the choice of path (more precisely, the path-ordered exponential depends only on the homotopy class of the path). We consider a path $\gamma = \gamma_2\gamma_1$ which is the concatenation of two straight-line paths $\gamma_1$ and $\gamma_2$ defined as follows: $\gamma_1$ is the line segment in the $(\tau_1,\tau_2)$-space from $(i\infty,i\infty)$ to a generic point $(\tau_1,i\infty)$ on the line $\tau_2=i\infty$, and $\gamma_2$ is the line segment from the point $(\tau_1,i\infty)$ to a generic point $(\tau_1,\tau_2)$. We may use the path-composition formula for the path-ordered exponential to write
\beq
\mathbb{P}_{\gamma_2\gamma_1}(\bs{\tau},\eps) = \mathbb{P}_{\gamma_2}(\bs{\tau},\eps)\,\mathbb{P}_{\gamma_1}(\bs{\tau},\eps)\,.
\eeq
We now discuss the contribution from each line segment in turn.

Let us start from the first segment. 
On $\gamma_1$, we need to solve the equation,
\begin{equation}
\partial_{\tau_1}\bs{M} = \eps\bs{\widetilde{A}}_1(\tau_1)\bs{M}\,, 
\end{equation}
where the matrix $\bs{\widetilde{A}}_1(\tau_1)$ is obtained from eq.~\eqref{eq:A_tau} by pulling it back to the path $\gamma_1$
\begin{equation}
\label{eq:canonicaltau2}
\rd\tau_1\,\bs{\widetilde{A}}_1(\tau_1) = \gamma_1^*\bA(\bs{\tau})\,.
\end{equation}
We find
\begin{equation}
\bs{\widetilde{A}}=\left(
\begin{array}{ccccccc}
 \tiny{\frac{5}{12} h_2^a+\frac{1}{2}h_2^b}& 0 & 0 & 0 & 0 & 0 & 0 \\
 0 & h_2^a  & 0 & 0 & 0 & 0 & 0 \\
 0 & 0 & h_2^b & 0 & 1 & 0 & 0 \\
 0 & 0 & 0 & h_2^b & 0 & 0 & 1 \\
 -21 \,h_3 & -3\,h_3 & h_4^b & 36\,h_4^a & h_2^b & 18\, h_3 & 0 \\
 \tiny{\frac{1}{2}h_2^b-\frac{11}{12}h_2^a}  & -\frac{1}{9}h_2^a  & \frac{1}{6}h_3 & h_3 & 0 & \frac{4}{3}h_2^a   & 0 \\
 -\frac{5}{2}  h_3 & -\frac{1}{2}h_3 & h_4^a & h_4^b & 0 & 3\,h_3 & h_2^b \\
\end{array}
\right)\,,
\end{equation}
with
\begin{align}
    h_2^a(\tau_1)=& -\frac{1}{4} ( u_1-3)^2 \Psi_1(\tau_1)^2\,,\nonumber\\
 h_2^b(\tau_1)=& \frac{9 - 108 u_1 + 6 u_1^2 - 12 u_1^3 - 7 u_1^4 }{24 (1 + u_1)^2}\Psi_1(\tau_1)^2\,,\nonumber\\
 h_3(\tau_1)=& \frac{(u_1-3)^2 ( u_1-1) }{4 \sqrt{3} (1 + u_1)}\Psi_1(\tau_1)^3\,,\\
 h_4^a(\tau_1)=&\frac{( u_1-1) (9 + 27 u_1 + 3 u_1^2 + u_1^3) (9 - 9 u_1 + 3 u_1^2 + 
   5 u_1^3) }{288 (1 + u_1)^4}\Psi_1(\tau_1)^4\,,\nonumber\\
 h_4^b(\tau_1)=& \frac{u_1^8+ 84 u_1^7+ 
   360 u_1^6+ 2052 u_1^5- 270 u_1^4- 5508 u_1^3+ 6048 u_1^2 - 1620 u_1  -891}{576 (1 + u_1)^4}\Psi_1(\tau_1)^4\,.\nonumber
\end{align}
We notice that $h_3$ is a modular form of weight 3 for the congruence subgroup $\Gamma^1(12)$, while $h_k^x$ ($k=2,4$ and $x=a,b$) are modular forms of weight $k$ (see appendix~\ref{app:modular} for details) for the same congruence subgroup. 
We also observe that no $\eps$-function appears in the differential equation. 
The first few orders in the $\eps$-expansion of the integrals read:
\begin{align}
    M_3(\tau_1,i\infty)=& \eps^3 \biggl[16\,\zeta_3+\frac{7}{4}   I(1,h_3,h_2^a;\tau_1)-\frac{3}{2} 
    I(h_3,h_2^b;\tau_1)\biggr]+\mathcal{O}(\eps^4)\,,\\
    M_4(\tau_1,i\infty)=&\eps^2 I(1,h_3;\tau_1) + \eps^3 \biggl[\frac{26}{3} \zeta_3 + 
   2 \zeta_2\log{\tilde{q}_1}+I(1,h_2^b,h_3;\tau_1) + 
   \frac{17}{24} I(1,h_3,h_2^a;\tau_1)\nonumber \\
   &+ 
   \frac{1}{4} I(1,h_3,h_2^b;\tau_1) + 
    I(h_2^b,1,h_3;\tau_1) \biggr]+\mathcal{O}(\eps^4)\,,\\
    M_5(\tau_1,i\infty)=&\eps^2   \biggl[\frac{7}{4} 
      I(h_3,h_2^a;\tau_1) - \frac{3}{2}
     I(h_3,h_2^b;\tau_1)\biggr] + 
 \eps^3 \biggl[
    28 \zeta_3+\frac{7}{4} I(h_2^b,h_3,h_2^a;\tau_1)\nonumber\\
    &- 
    \frac{3}{2} I(h_2^b,h_3,h_2^b ;\tau_1) + 
    \frac{119}{48} I(h_3,h_2^a,h_2^a ;\tau_1) - 
    \frac{5}{8} I(h_3,h_2^a,h_2^b ;\tau_1) \nonumber\\
    &- 
    \frac{5}{8} I(h_3,h_2^b,h_2^a ;\tau_1) - 
    \frac{3}{4} I(h_3,h_2^b,h_2^b ;\tau_1) + 
    36 I(h_4^a,1,h_3 ;\tau_1) \biggr]+\mathcal{O}(\eps^4)\,,\\
    M_6(\tau_1,i\infty)=&\frac{4}{3} + \eps \biggl[\frac{3}{4} I(h_2^a;\tau_1) + 
    \frac{1}{2} I(h_2^b;\tau_1)\biggr] + 
  \eps^2 \biggl[2\,\zeta_2+\frac{73}{144} 
      I(h_2^a,h_2^a ;\tau_1) + 
    \frac{5}{24} I(h_2^a,h_2^b ;\tau_1) \nonumber\\
    &+ 
    \frac{5}{24} I(h_2^b,h_2^a ;\tau_1) + 
    \frac{1}{4} I(h_2^b,h_2^b ;\tau_1)  
     \biggr] + \eps^3\biggl[ 
    \frac{9}{8}\zeta_2 I(h_2^a;\tau_1)   + 
    \frac{3}{4}\zeta_2 I(h_2^b;\tau_1)   - 
    4 \zeta_3 \nonumber\\
    &+\frac{701}{1728} I(h_2^a,h_2^a, h_2^a;\tau_1) + 
    \frac{25}{288} I(h_2^a,h_2^a,h_2^b ;\tau_1) + 
    \frac{25}{288} I(h_2^a,h_2^b,h_2^a ;\tau_1) \nonumber\\
    &+ 
    \frac{5}{48} I(h_2^a,h_2^b,h_2^b ;\tau_1) + 
    \frac{25}{288} I(h_2^b,h_2^a,h_2^a ;\tau_1) + 
    \frac{5}{48} I(h_2^b,h_2^a,h_2^b;\tau_1) \nonumber\\
    &+ 
    \frac{5}{48} I(h_2^b,h_2^b,h_2^a ;\tau_1) + 
    \frac{1}{8} I(h_2^b, h_2^b, h_2^b;\tau_1) + 
     I(h_3,1,h_3 ;\tau_1) \biggr]+\mathcal{O}(\eps^4)\,,\\
     M_7(\tau_1,i\infty)=&\eps\, I(h_3;\tau_1)+ 
  \eps^2\biggl[2 \zeta_2+  
      I(h_2^b,h_3;\tau_1) + 
    \frac{17}{24} I(h_3,h_2^a ;\tau_1) + 
    \frac{1}{4} I(h_3,h_2^b ;\tau_1) \biggr] \nonumber\\
    &+ 
  \eps^3 \biggl[\frac{35}{3} \zeta_3+ 
    2\zeta_2 \,I(h_2^b;\tau_1)  + 
    \frac{3}{2}\zeta_2 \,I(h_3;\tau_1)  + I(h_2^b,h_2^b,h_3;\tau_1)  \nonumber\\
    &+ 
    \frac{17}{24} I(h_2^b,h_3,h_2^a ;\tau_1) + 
    \frac{1}{4} I(h_2^b,h_3,h_2^b ;\tau_1) + 
    \frac{169}{288} I(h_3,h_2^a,h_2^a ;\tau_1)  \nonumber\\
    &+ 
    \frac{5}{48} I(h_3,h_2^a,h_2^b;\tau_1) + 
    \frac{5}{48} I(h_3,h_2^b,h_2^a ;\tau_1) + 
    \frac{1}{8} I(h_3,h_2^b,h_2^b;\tau_1)  \nonumber\\
    &+ 
     I(h_4^b,1,h_3;\tau_1) 
    \biggr]+\mathcal{O}(\eps^4)\,.
\end{align}
%Substituting the periods by their series expansion, it is possible to check that our solution has uniform weight.

Next we integrate along $\gamma_2$. We then need to solve the equation
\begin{equation}
\partial_{\tau_2}\bs{M} = \eps\bs{{A}}_2(\tau_2)\bs{M}\,, 
\end{equation}
where the matrix $\bs{{A}}_2$ is given in eq.~\eqref{mat:canonicalforms}.
Note that $\bs{{A}}_2(\tau_2)$ still depends on $\tau_1$, but we drop this dependence from the argument, because $\tau_1$ is a constant on the line segment $\gamma_2$.
%Eq.~\eqref{eq:canonicaltau2} becomes
%\begin{equation}
%    \bA(\tau_2)=\bA_2(\tau_2)_{|\tau_1}\mathrm{d}\tau_2\,,
%\end{equation}
%where $\bA_2(\tau_2)_{|\tau_1}$ is the same as in Eq.~\eqref{mat:canonicalforms}, with the difference that $\tau_1$ does not vary. 
Since the integration boundaries are different from $i\infty$, $\eps$-functions appear in the result. Putting everything together, we find for the leading term in the $\eps$-expansion for $M_3(\tau_1,\tau_2)$ is
\begin{align}
    M_3(\tau_1,\tau_2)=&\eps^3\biggl[16 \,\zeta_3+ 
    \log{\tilde{q}_2}\,I(1,h_3;\tau_1) +\frac{7}{4} I(1,h_3,h_2^a;\tau_1) - 
   \frac{3}{2} I(1,h_3,h_2^b;\tau_1) \nonumber\\
 \label{eq:M3_result}  &- 
   \frac{7}{4} I(h_2^a;\tau_1) I(1,\phi_{-1,3};\tau_2) + 
   \frac{3}{2} I(h_2^b;\tau_1) I(1,\phi_{-1,3};\tau_2) \\
   &- 
    I(1,\phi_{-1,3},\phi_{0,2}^b;\tau_2) - 
    I(1,\phi_{-1,3}, \phi_{0,2}^a ;\tau_2) 
    \biggr]+\mathcal{O}(\eps^4)\,.\nonumber
\end{align}
The $\eps$-expansion for the all the integrals up to $\mathcal{O}(\eps^5)$ can be found in an ancillary file. 
As a check of our computation, we choose a point that lies within the radius of convergence of the series expansion of the holomorphic period $\psi_0$ near the MUM-point $\bs{z}=0$ with negative momentum squared. We expanded all the iterated integrals as series up to order 18.  We verified numerically against AMFlow~\cite{Liu:2022chg} up to $\mathcal{O}(\eps^4)$ and found perfect agreement within 16 digits.

Let us conclude by making a comment about the leading term in the $\eps$-expansion of $M_3$ shown in eq.~\eqref{eq:M3_result}, which corresponds to the master integral evaluated in strictly $D=2$ dimensions. We observe that the iterated integrals only involve the integration kernels $h_k$, $h_k^x$ and $\phi^k_{i,j}$, but no $\eps$-functions $\mathbb{I}_k$ appear. We have already seen $h_k$ and $h_k^x$ are modular forms for $\Gamma^1(12)$. It is easy to see (cf. appendix~\ref{app:modular}) that for $\tau_1$ fixed also the $\phi^k_{i,j}$ define meromorphic modular forms for $\Gamma_1(12)$. We conclude that in $D=2$ dimensions, the master integral $M_3$ can be expressed in terms of genuine iterated integrals of meromorphic modular forms, and no $\eps$-functions appear. This is similar to what happens for the equal-mass three-loop integral~\cite{Broedel:2019kmn}. 
%Since the integrals $M_4$, $M_5$ and $M_7$ are obtained from $M_3$ by differentiation, the same conclusion holds for those integrals (the integral $I_6$ vanishes in $D=2$ dimensions \claudecomment{correct?}\sm{For $M_6$ I get 4/3 at $\eps^0$, one can see that it is non zero also from the equal mass banana, were IBP reducing $I_6$ one gets $ 4/3 I_1 + 1/6  I_3$. Regarding $M_4$, $M_5$ it is true that there is no $\eps$-function, but it is not true for $M_7$ because it also gets contribution from the tadpoles and $M_6$}). This is similar to the situation encountered for the equal-mas case, where at $\eps=0$ iterated integrals of modular forms are sufficent ot express the answer~\cite{Broedel:2019kmn}.

\section{Conclusion}
\label{sec:conclusions}

In this paper we have presented for the first time fully analytic results for all the master integrals for the three-loop banana integrals with three equal masses in $D=2-2\eps$ dimensions. While analytic results for the three-loop banana integrals with four different masses have recently appeared~\cite{Duhr:2025kkq,Pogel:2025bca}, the main novelty of our result is that the master integrals can be expressed in terms of iterated integrals of meromorphic modular forms (and integrals thereof). While meromorphic modular forms have already appeared in the context of the equal-mass banana integral~\cite{Broedel:2019kmn,Broedel:2021zij,Pogel:2022yat}, this is the first time that it was possible to express a Feynman integral depending on two dimensionless ratios in terms of this class of functions.

The appearance of meromorphic modular forms has some practical advantages, because this class of functions is relatively well studied in the literature. In particular, we have used our computation to study the properties of our results. We were able to rigorously prove that our canonical differential equations satisfy the properties conjectured in ref.~\cite{Duhr:2025lbz}, and we could show that the differential forms in the canonical differential equation have only simple poles and define independent cohomology classes. We stress that all our results rely very profoundly on various recent advances in our understanding of the mathematics underlying Feynman integrals associated with K3 surfaces, in particular their period geometry~\cite{Duhr:2025tdf,Duhr:2025ppd}, the corresponding canonical bases~\cite{Gorges:2023zgv,Pogel:2022ken,Pogel:2022vat,Pogel:2022yat,Pogel:2025bca,Duhr:2025lbz,Maggio:2025jel,e-collaboration:2025frv,Bree:2025tug} and their properties~\cite{Pogel:2024sdi,Duhr:2024xsy,Duhr:2025xyy}, and the mathematics underlying iterated integrals of meromorphic modular forms~\cite{ManinModular,Brown:2014pnb,Matthes2022IteratedPrimitives,Broedel:2021zij,Brown:2025zsw}.

For the future, it would be interesting to see if the approach developed here to solve Feynman integrals associated with K3 surfaces that are products of two elliptic curves can be applied also to other cases. For example, in ref.~\cite{Dlapa:2024cje} it was observed that a certain family of integrals appearing in the context of gravitational wave scattering involves a K3 surface whose periods can be written as products of complete elliptic integrals. We expect that it is possible to express those results in terms of iterated integrals of modular forms as well.

\subsection*{Acknowledgments}
We are grateful to Matthias Carosi and Federico Gasparotto for discussions, and to Franziska Porkert, Cathrin Semper, Yoann S\"ohnle and Sven Stawinski for collaborations on related projects.
This work was funded by the European Union (ERC Consolidator Grant LoCoMotive 101043686). Views and opinions expressed are however those of the author(s) only and do not necessarily reflect those of the European Union or the European Research Council. Neither the European Union nor the granting authority can be held responsible for them.

\begin{appendix}

% !TEX root = main.tex

\section{Review of modular forms}
\label{app:modular}

\subsection{Modular forms in one variable}
Let $\Gamma$ be a subgroup of $\SL_2(\mathbb{Z})$ (of finite index). It acts on the complex upper half-plane $\mathbb{H} = \big\{\tau\in\mathbb{C}:\textrm{Im}\,\tau>0\big\}$ via M\"obius transformations,
\beq
\gamma\cdot \tau = \tfrac{a\tau+b}{c\tau+d}\,,\qquad \gamma=\left(\begin{smallmatrix}a&b\\c&d\end{smallmatrix}\right)\in\Gamma\,.
\eeq
A meromorphic modular form of weight $k$ is a meromorphic function $f:\mathbb{H}\to \mathbb{C}$ such that
\beq
f(\gamma\cdot \tau) = (c\tau+d)^k\,f(\tau)\,,\qquad \textrm{for all }\gamma\in\Gamma\,.
\eeq
A modular function for $\Gamma$ is a meromorphic modular form of weight 0.
We denote by $\cM_k(\Gamma)$ the vector space generated by all meromorphic modular forms for $\Gamma$, and $M_k(\Gamma)$ is the subspace by all (holomorphic) modular forms. 
%We also write $\cM(\Gamma) = \bigoplus_k\cM_k(\Gamma)$. 
Note that $M_k(\Gamma)$ is non-trivial only for $k>0$.

We will only be interested in cases where $\Gamma$ has genus zero, by which we mean that the Riemann surface $Y_{\Gamma} = \Gamma\setminus \mathbb{H}$ has genus zero. The field of meromorphic functions of $Y_{\Gamma}$ can be identified with the field of modular functions $\cM_0(\Gamma)$. Since $Y_{\Gamma}$ has genus zero, its field of meromorphic functions has a single generator, called a \emph{Hauptmodul} for $\Gamma$. Said differently, every modular function can be expressed as a rational function in the Hauptmodul.

Throughout this paper, we encountered  modular forms for the congruence subgroups $\Gamma_1(6)$ and $\Gamma_1(12)$.\footnote{Modular forms for $\Gamma^1(N)$ can be related to those for $\Gamma_1(N)$, cf. ref.~\cite{Duhr:2025ppd}.} In ref.~\cite{Broedel:2018rwm} it was shown that any modular form $f$ of weight $k$ for $\Gamma_1(6)$, i.e., any element of $M_k(\Gamma_1(6))$ can
be cast in the form
\beq\label{eq:f_G16_decomp}
f(\tau)=\psi(\tau)^k\,\sum_{r=0}^kc_k\,t(\tau)^k\,,
\eeq 
where the $c_k$ are constants, $\psi$ is the modular form of weight one for $\Gamma_1(6)$ defined in eq.~\eqref{eq:psi_def} and $t$ is the Hauptmodul in eq.~\eqref{eq:G06_Hauptmodul}.\footnote{$\Gamma_0(6)$ and $\Gamma_1(6)$ admit the same Hauptmodul, because, even though $\Gamma_0(6)$ and $\Gamma_1(6)$ are different as subgroups of $\SL_2(\mathbb{Z})$, they become identical as subgroups in PSL$_2(\mathbb{Z})$, so that the Riemann surfaces $Y_{\Gamma_0(6)}$ and $Y_{\Gamma_1(6)}$ are isomorphic.} We now provide a similar description of modular forms for $\Gamma_1(12)$. Since $\Gamma_1(12)\subset \Gamma_1(6)$, we have $M_k(\Gamma_1(6))\subset M_k(\Gamma_1(12))$, and so $\psi$ is also a modular form of weight one for $\Gamma_1(12)$. Let $f\in M_k(\Gamma_1(12))$. From eq.~\eqref{eq:t_to_u} and eq.~\eqref{eq:f_G16_decomp}, we obtain
\beq
\frac{f(\tau)}{\psi(\tau)^k}=\sum_{r=0}^kc_k\,\left(\frac{u(\tau)(3-u(\tau))}{1+u(\tau)}\right)^r\,.
\eeq
The right-hand side has a pole at $u(\tau)=-1$. Since $f$ is holomorphic everywhere, this pole must come from a zero of $\psi$. We can easily check that this pole is simple. Following exactly the same argument as in the derivation of eq.~\eqref{eq:f_G16_decomp} in ref.~\cite{Broedel:2018rwm}, we find that every modular form $f\in M_k(\Gamma_1(12))$ can be cast in the form
\beq
f(\tau) = \frac{\psi(\tau)^k}{(1+u(\tau))^k}\sum_{r=0}^{2k}c_k\,u(\tau)^k\,.
\eeq

In the context of canonical differential equations satisfied by Feynman integrals, we are typically interested in differential forms of the type $\rd\tau\,f(\tau)$, where $f$ is a meromorphic modular form for $\Gamma$ (of positive weight). 
It is possible to identify a finite set of such differential forms that define independent cohomology classes, i.e., a finite set such that every differential form $\rd\tau\,f(\tau)$ is a linear combination from this set, up to a total derivative~\cite{Matthes2022IteratedPrimitives,Broedel:2021zij,Brown:2025zsw}. In order to state the result, it is useful to consider a slightly larger class of functions. A meromorphic quasi-modular form of weight $k$ and depth $p$ for $\Gamma$ is a meromorphic function $f:\mathbb{H}\to \mathbb{C}$ such that
\beq
f( \gamma\cdot \tau) = (c\tau+d)^k\,\sum_{r=0}^pf_r(\tau)\,\left(\frac{c}{c\tau +d}\right)^r\,,
\eeq
where $f_0,\ldots,f_p$ are meromorphic functions with $f_p\neq0$. We denote the vector space of all quasi-modular forms of weight $k$ (of any depth) for $\Gamma$ by $\mathcal{QM}_k(\Gamma)$. Note that quasi-modular forms of depth zero are precisely the modular forms, so that $\cM_k(\Gamma)\subseteq\mathcal{QM}_k(\Gamma)$. Unlike $\cM_k(\Gamma)$, however, $\mathcal{QM}_k(\Gamma)$ is closed under differentiation (which raises the depth by one unit).
We have~\cite{Matthes2022IteratedPrimitives,Broedel:2021zij}
\beq\label{eq:QM_decomp}
\mathcal{QM}_k(\Gamma) =\left\{\begin{array}{ll}
\cM_{k}(\Gamma)\oplus \partial_{\tau}\mathcal{QM}_{k-2}(\Gamma)\,, & k <2\,,\\
\widetilde{\cM}_k(\Gamma)\oplus \cM_{2-k}(\Gamma)G_2^{k-1}\oplus\partial_{\tau}\mathcal{QM}_{k-2}(\Gamma)\,,& k\ge 2\,,
\end{array}\right.
\eeq
where $G_2$ is the Eisenstein series of weight two for $\SL_2(\mathbb{Z})$ and $\widetilde{\cM}_k(\Gamma)$ is a finite-dimensional vector space. For the precise definition of $\widetilde{\cM}_k(\Gamma)$ we refer to refs.~\cite{Matthes2022IteratedPrimitives,Broedel:2021zij}. Here it suffices to say that all elements in $\widetilde{\cM}_k(\Gamma)$ have pole orders bounded by $k-1$ at points $\tau_0\in\mathbb{H}$. The notation $\widetilde{\cM}(\Gamma) = \bigoplus_k\widetilde{\cM}_k(\Gamma)$ will be useful later on. Equation~\eqref{eq:QM_decomp} implies that every meromorphic modular form $f\in\cM_k(\Gamma)$ with $k\ge 2$ can be written as $f=\tilde{f}+\partial_{\tau}g$ with $\tilde{f}\in\widetilde{\cM}_k(\Gamma)$ and $g\in \mathcal{QM}_{k-2}(\Gamma)$. Note that we have
\beq\label{eq:proof_eq_2}
\widetilde{\cM}(\Gamma)\cap \partial_{\tau}\mathcal{QM}(\Gamma)=\{0\}\,.
\eeq

\subsection{Modular forms in two variables}
We now consider two subgroups $\Gamma$ and $\Gamma'$ of $\SL_2(\mathbb{Z})$ (each of finite index). A modular form in two variables of weights $(k_1,k_2)$ for $\Gamma\times \Gamma'$ is  a meromorphic function $f:\mathbb{H}^2\to\mathbb{C}$ such that for all $(\gamma_1,\gamma_2)\in\Gamma\times\Gamma'$ we have
\beq
f(\gamma_1\cdot \tau_1,\gamma_2\cdot\tau_2) = (c_1\tau_1+d_1)^{k_1}\,(c_2\tau_2+d_2)^{k_2}\,f(\tau_1,\tau_2)\,.
\eeq
We denote by $\cM_{k_1,k_2}(\Gamma,\Gamma')$ the vector space of all meromorphic modular forms of weights $(k_1,k_2)$ for $\Gamma\times\Gamma'$. It is easy to check that (see, e.g., ref.~\cite{Duhr:2025ppd})
\beq
\cM_{k_1,k_2}(\Gamma,\Gamma') = \cM_{k_1}(\Gamma)\otimes \cM_{k_2}(\Gamma')\,.
\eeq
In other words, all modular forms in two variables for $\Gamma\times\Gamma'$ can be written as linear combinations of products of modular forms in one variable. We can similarly define the algebra of quasi-modular forms for $\Gamma\times\Gamma'$,
\beq
\mathcal{QM}(\Gamma,\Gamma')=\bigoplus_{k_1,k_2}\mathcal{QM}_{k_1}(\Gamma)\otimes\mathcal{QM}_{k_2}(\Gamma')\,.
\eeq

In the context of canonical differential equations for Feynman integrals, one is interested in differential one-forms of the form $\omega_{ij} = \rd\tau_1\,\omega_{ij}^1+\rd\tau_2\,\omega_{ij}^2$. We say that such a differential form transforms like a modular form of weights $(k_1,k_2)$ for $\Gamma\times\Gamma'$ if under modular transformations we have
\beq
\omega_{ij}\mapsto (c_1\tau_1+d_1)^{k_1}\,(c_2\tau_2+d_2)^{k_2}\,\omega_{ij}\,.
\eeq
This is equivalent to saying that $\omega_{ij}^1$ and $\omega_{ij}^2$ are modular forms in two variables for $\Gamma\times\Gamma'$ with weights $(k_1+2,k_2)$ and $(k_1,k_2+2)$, respectively.

\section{Differential forms in the canonical differential equation}
\label{app:diff_forms_can_eq}
The explicit expressions for the differential forms $\omega^k_{ij}$ depend on rational functions in $u_1,u_2$, holomorphic periods and the integrals defined in eq.~\eqref{new_int}. Here we collect the rational functions and the periods in 20 functions $\phi_{k,l}$  shown at the end of this appendix. The $\omega^k_{ij}$ read 
\begin{align}
\omega_{11}^1 =& 
 \frac{1}{3} (2 \phi_{2,0}^b-\phi_{2,0}^a  
    )\,,\nonumber\\ 
\omega_{22}^1 =& 
\phi_{2,0}^b\,,\nonumber\\  \omega_{33}^1 =& 
 \frac{1}{2} (2 \bbI_2 + \phi_{2,0}^b + \phi_{2,0}^c)\,,\nonumber\\ 
\omega_{41}^1 =& 
 21 \bbI_3\,,\nonumber\\ \omega_{42}^1 =& 
 3 \bbI_3\,,\nonumber\\ \omega_{43}^1 =& -2 \bbI_1 \bbI_2 - 9 
\bbI_3^2 + \phi_{2,2}\,,\nonumber\\ \omega_{44}^1 =& 
 \frac{1}{2} ( \phi_{2,0}^b + \phi_{2,0}^c-2 \bbI_2)\,,\nonumber\\ \omega_{45}^1 =& -2 \bbI_1\,,\nonumber\\ \omega_{46}^1 =& -18 \bbI_3\,,\nonumber\\  \omega_{51}^1 =& 
 21 \phi_{3,-1}\,,\nonumber\\ \omega_{52}^1 =& 
 3 \phi_{3,-1}\,,\nonumber\\ \omega_{53}^1 =& -
\bbI_2^2 - 
  18 \bbI_3 \phi_{3,-1} - 
  2 \bbI_1 \phi_{4,-2} + \phi_{4,0} - 
  \phi_{4,-2} (\phi_{0,2}^b + \phi_{0,2}^c) + 
  \frac{1}{4} (\phi_{2,0}^b + \phi_{2,0}^c)^2\,,\nonumber\\ \omega_{54}^1 =& -2 \phi_{4,-2}\,,\nonumber\\ \omega_{55}^1 =& 
 \frac{1}{2} ( \phi_{2,0}^b + \phi_{2,0}^c-2 \bbI_2)\,,\nonumber\\ \omega_{56}^1 =& -18 \phi_{3,-1}\,,\nonumber\\  \omega_{61}^1 =&
  -\frac{1}{3} (\phi_{2,0}^a + 
    2 \phi_{2,0}^b)\,,\nonumber\\ \omega_{62}^1 =& -\frac{1}{9}
    \phi_{2,0}^b\,,\nonumber\\ \omega_{63}^1 =& 
 \frac{1}{6} ( 2 \phi_{3,1}^a - 
    3 \phi_{3,-1} \phi_{0,2}^b + 
    5 \bbI_3 \phi_{2,0}^b - 
    4 \phi_{3,1}^b - 
    3 \phi_{3,-1} \phi_{0,2}^c - 
    3 \bbI_3 \phi_{2,0}^c-6 \bbI_2 \bbI_3 - 
    6 \bbI_1 \phi_{3,-1} )\,,\nonumber\\ \omega_{64}^1 =& -\phi_{3,-1}\,,\nonumber\\ \omega_{65}^1 =& -\bbI_2\,,\nonumber\\ \omega_{66}^1 =&\frac{4}{3} 
  \phi_{2,0}^b\,,\nonumber\\  \omega_{71}^1 =& 
 \frac{1}{2} (42 \bbI_2 \bbI_3 + 
    42 \bbI_1 \phi_{3,-1} + 
    2 \bbI_3 \phi_{2,0}^a - 
    20 \phi_{3,1}^a + 
    21 \phi_{3,-1} \phi_{0,2}^b - 
    31 \bbI_3 \phi_{2,0}^b + 
    32 \phi_{3,1}^b \nonumber\\
    &+ 
    21 \phi_{3,-1} \phi_{0,2}^c + 
    21 \bbI_3 \phi_{2,0}^c)\,,\nonumber\\ \omega_{72}^1 =& 
 3 \bbI_2 \bbI_3 + 
  3 \bbI_1 \phi_{3,-1} + \phi_{3,1}^b + 
  \frac{3}{2} \phi_{3,1} (\phi_{0,2}^b + \phi_{0,2}^c) - 
  \frac{1}{2} \bbI_3 (7 \phi_{2,0}^b - 
     3 \phi_{2,0}^c)\,,\nonumber\\ \omega_{73}^1 =& 
 2 \bbI_2 ( \phi_{2,2}-9 
\bbI_3^2) - 2 
\bbI_1^2 \phi_{4,-2} + 
  \phi_{4,2} + 
  \phi_{4,0} (\phi_{0,2}^b + \phi_{0,2}^c) - 
  \frac{1}{2} \phi_{4,-2} (\phi_{0,2}^b + \phi_{0,2}^c)^2 \nonumber\\
  &+ 
  6 \bbI_3 (2 \phi_{3,1}^a - 4 \phi_{3,1}^b - 
     3 \phi_{3,-1} (\phi_{0,2}^b + \phi_{0,2}^c)) + 3 
\bbI_3^2 (5 \phi_{2,0}^b - 3 \phi_{2,0}^c) + 
  \phi_{2,2} (\phi_{2,0}^b + \phi_{2,0}^c) \nonumber\\
  &+ 
  2 \bbI_1 (-
\bbI_2^2 - 
     18 \bbI_3 \phi_{3,-1} + \phi_{4,0} - 
     \phi_{4,-2} (\phi_{0,2}^b + \phi_{0,2}^c) + 
     \frac{1}{4} (\phi_{2,0}^b + \phi_{2,0}^c)^2)\,,\nonumber\\ \omega_{74}^1 =&  \phi_{4,0} - 
  \phi_{4,-2} (\phi_{0,2}^b + \phi_{0,2}^c) + 
  \frac{1}{4} (\phi_{2,0}^b + \phi_{2,0}^c)^2-
\bbI_2^2 - 
  18 \bbI_3 \phi_{3,-1} - 
  2 \bbI_1 \phi_{4,-2} \,,\nonumber\\ \omega_{75}^1 =& -2 \bbI_1
    \bbI_2 - 9 
\bbI_3^2 + \phi_{2,2}\,,\nonumber\\ \omega_{76}^1 =& -3 (6 \bbI_2 \bbI_3 + 6 \bbI_1 \phi_{3,-1} - 
    2 \phi_{3,1}^a + 
    3 \phi_{3,-1} \phi_{0,2}^b - 
    5 \bbI_3 \phi_{2,0}^b + 
    4 \phi_{3,1}^b + 
    3 \phi_{3,-1} \phi_{0,2}^c + 
    3 \bbI_3 \phi_{2,0}^c)\,,\nonumber\\ \omega_{77}^1 =& 
 \frac{1}{2} (2 \bbI_2 + \phi_{2,0}^b + \phi_{2,0}^c
    )\,.
\end{align}
\begin{align}
\omega^2_{11} =&
 \frac{1}{3} (2 \phi_{0,2}^b-\phi_{0,2}^a )\,,\nonumber\\  \omega^2_{22} =&\
\phi_{0,2}^b\,,\nonumber\\ \omega^2_{33} =&
 \bbI_1 + 
  \frac{1}{2} (\phi_{0,2}^b + \phi_{0,2}^c)\,,\nonumber\\ 
\omega^2_{41} =&
 21 \phi_{-1,3}\,,\nonumber\\ \omega^2_{42} =&
 3 \phi_{-1,3}\,,\nonumber\\ \omega^2_{43} =&  \phi_{0,4} + 
  \frac{1}{4} (\phi_{0,2}^b + \phi_{0,2}^c)^2 - 
  \phi_{-2,4} (\phi_{2,0}^b + \phi_{2,0}^c)-
\bbI_1^2 - 
  2 \bbI_2 \phi_{-2,4} - 
  18 \bbI_3 \phi_{-1,3}\,,\nonumber\\ \omega^2_{44} =&
 \frac{1}{2} ( \phi_{0,2}^b + \phi_{0,2}^c-2 \bbI_1 )\,,\nonumber\\ \omega^2_{45} =&-2 \phi_{-2,4}\,,\nonumber\\ \omega^2_{46} =&-18 \phi_{-1,3}\,,\nonumber\\  \omega^2_{51} =&
 21 \bbI_3\,,\nonumber\\ \omega^2_{52} =&
 3 \bbI_3\,,\nonumber\\ \omega^2_{53} =&-2 \bbI_1 \bbI_2 - 9 
\bbI_3^2 + \phi_{2,2}\,,\nonumber\\ \omega^2_{54} =&-2 \bbI_2\,,\nonumber\\ \omega^2_{55} =&
 \frac{1}{2} ( \phi_{0,2}^b + \phi_{0,2}^c-2 \bbI_1)\,,\nonumber\\ \omega^2_{56} =&-18 \bbI_3\,,\nonumber\\  \omega^2_{61} =&
 -\frac{1}{3} (\phi_{0,2}^a + 
    2 \phi_{0,2}^b)\,,\nonumber\\ \omega^2_{62} =&-\frac{1}{9}
    \phi_{0,2}^b\,,\nonumber\\ \omega^2_{63} =&
 \frac{1}{6} ( 2 \phi_{1,3}^a -6 \bbI_1 \bbI_3 - 
    6 \bbI_2 \phi_{-1,3} + 
    5 \bbI_3 \phi_{0,2}^b - 
    4 \phi_{1,3}^b - 
    3 \phi_{-1,3} \phi_{2,0}^b - 
    3 \bbI_3 \phi_{0,2}^c - 
    3 \phi_{-1,3} \phi_{2,0}^c)\,,\nonumber\\ \omega^2_{64} =&-\bbI_3\,,\nonumber\\ \omega^2_{65} =&-\phi_{-1,3}\,,\nonumber\\ \omega^2_{66} =&
 \frac{4}{3} \phi_{0,2}^b\,,\nonumber\\ \omega^2_{71} =&
 \frac{1}{2} (42 \bbI_1 \bbI_3 + 
    42 \bbI_2 \phi_{-1,3} + 
    2 \bbI_3 \phi_{0,2}^a - 
    20 \phi_{1,3}^a - 
    31 \bbI_3 \phi_{0,2}^b + 
    32 \phi_{1,3}^b + 
    21 \phi_{-1,3} \phi_{2,0}^b \nonumber\\
    &+ 
    21 \bbI_3 \phi_{0,2}^c + 
    21 \phi_{-1,3} \phi_{2,0}^c)\,,\nonumber\\ \omega^2_{72} =&
 3 \bbI_1 \bbI_3 + 
  3 \bbI_2 \phi_{-1,3} + \phi_{1,3}^b - 
  \frac{1}{2} \bbI_3 (7 \phi_{0,2}^b - 3 \phi_{0,2}^c) + 
  \frac{3}{2} \phi_{-1,3}^b (\phi_{2,0}^b + \phi_{0,2}^c)\,,\nonumber\\ \omega^2_{73} =& 
  \phi_{0,4} (\phi_{2,0}^b + \phi_{2,0}^c) - 
  \frac{1}{2} \phi_{-2,4} (\phi_{2,0}^b + \phi_{2,0}^c)^2+ \phi_{2,4}+ 
  \phi_{2,2} (\phi_{0,2}^b + \phi_{0,2}^c)-2 
\bbI_1^2 \bbI_2 - 2 
\bbI_2^2 \phi_{-2,4} \nonumber\\
    &- 
  36 \bbI_2 \bbI_3 \phi_{-1,3} + 
  2 \bbI_1 (\phi_{2,2}-9 
\bbI_3^2 )  + 24 
\bbI_3^2 \phi_{0,2}^b + 12 \bbI_3 (\phi_{1,3}^a - 2 \phi_{1,3}^b) - 9 
\bbI_3^2 (\phi_{0,2}^b + \phi_{0,2}^c) \nonumber\\
    & - 
  18 \bbI_3 \phi_{-1,3} (\phi_{2,0}^b + \phi_{2,0}^c)  + 
  2 \bbI_2 (\phi_{0,4} + 
     \frac{1}{4} (\phi_{0,2}^b + \phi_{0,2}^c)^2 - 
     \phi_{-2,4} (\phi_{2,0}^b + \phi_{2,0}^c))\,,\nonumber\\ \omega^2_{74} =& \phi_{2,2}-2 \bbI_1
    \bbI_2 - 9 
\bbI_3^2\,,\nonumber\\ \omega^2_{75} =& \phi_{0,4} + 
  \frac{1}{4} (\phi_{0,2}^b + \phi_{0,2}^c)^2 - 
  \phi_{-2,4} (\phi_{2,0}^b + \phi_{2,0}^c)-
\bbI_1^2 - 
  2 \bbI_2 \phi_{-2,4} - 
  18 \bbI_3 \phi_{-1,3}\,,\nonumber\\ \omega^2_{76} =&-3 (6 \bbI_1
      \bbI_3 + 6 \bbI_2 \phi_{-1,3} - 
    2 \phi_{1,3}^a - 
    5 \bbI_3 \phi_{0,2}^b + 
    4 \phi_{1,3}^b + 
    3 \phi_{-1,3} \phi_{2,0}^b + 
    3 \bbI_3 \phi_{0,2}^c + 
    3 \phi_{-1,3} \phi_{2,0}^c)\,,\nonumber\\ \omega^2_{77} =&
 \bbI_1 + \frac{1}{2} (\phi_{0,2}^b + \phi_{0,2}^c).
\end{align}

\begin{align}
%h(\bs{\tau})=&36 (1 + u_2)^3 (u_1^2 + u_2^2 - 
        % u_1 ( u_2^2-3)) (12 u_1 (3 + u_2^2)^2 - %4 u_1^3 (3 + u_2^2)^2 \nonumber\\
         %&+ 
         %9 (9 - 10 u_2^2 + u_2^4) + u_1^4 (9 - 10 u_2^2 + u_2^4) - 
         %2 u_1^2 (9 + 54 u_2^2 + u_2^4)) \,, \nonumber\\
\phi_{4,-2}(\bs{\tau}) =&\frac{3 \Psi_1(\tau_1)^4}{4  L_1(\bs{u})L_2(\bs{u}) (1 + u_1)^2 \Psi_1(\tau_2)^2}  ( u_1-3)^2 (  u_1-1) u_1 (3 + u_1) (1 + u_2)^2 \nonumber\\
&(-81 - 162 u_1 + 189 u_1^2 - 
    108 u_1^3 - 135 u_1^4 - 18 u_1^5 - 5 u_1^6 + 45 u_2^2 - 
    54 u_1 u_2^2 \nonumber\\
&+ 135 u_1^2 u_2^2 - 36 u_1^3 u_2^2 - 
    21 u_1^4 u_2^2 - 6 u_1^5 u_2^2 + 
    u_1^6 u_2^2)\,, \nonumber\\
\phi_{3,-1}(\bs{\tau}) =& -\frac{(-3 + u_1)^2 (-1 + u_1) u_1 (3 + u_1) (1 + u_2) \Psi_1(\tau_1)^3}{
 6 (1 + u_1) (u_1^2 + u_2^2 - u_1 (-3 + u_2^2)) \Psi_1(\tau_2)  }  \,, \nonumber\\   
   \phi_{2,0}^a(\bs{\tau}) =&\frac{(3 + u_1^2) ( u_1^2+ 6 u_1-3) }{4 (1 + u_1)^2}\Psi_1(\tau_1)^2\,, \nonumber\\
 \phi_{2,0}^b(\bs{\tau}) =&-\frac{(-3 + u_1)^2 (u_1^2 - u_2^2 + u_1 (3 + u_2^2)) \Psi_1(\tau_1)^2}{4 (u_1^2 + u_2^2 - u_1 (-3 + u_2^2))}\,, \nonumber\\
 \phi_{2,0}^c(\bs{\tau}) =&-\frac{12\,\Psi_1(\tau_1)^2}{36  L_1(\bs{u})L_2(\bs{u})(1 + u_1)^2}  (u_1^2 + u_2^2 - u_1 ( u_2^2-3))(3 + u_1^2) (-3 + 6 u_1 + u_1^2) (9 \nonumber\\
 &+ 6 u_1 - 3 u_1^2 - 
      6 u_2 - 2 u_1^2 u_2 - 3 u_2^2 - 2 u_1 u_2^2 + 
      u_1^2 u_2^2) (9 + 6 u_1 - 3 u_1^2 + 6 u_2 \nonumber\\
 &+ 
      2 u_1^2 u_2 - 3 u_2^2 - 2 u_1 u_2^2 + 
      u_1^2 u_2^2)  \,, \nonumber\\    
\phi_{4,0}(\bs{\tau}) =&-\frac{\Psi_1(\tau_1)^4}{24 (1 + u_1)^2 (u_1^2 + u_2^2 - u_1 (-3 + u_2^2))^2}(-3 + 
   u_1)^2 (2 u_1^8 + 18 u_2^4 + u_1^7 (21 - 5 u_2^2) \nonumber\\
 &- 
   9 u_1 u_2^2 (-15 + 7 u_2^2) + u_1^5 (198 - 31 u_2^2 + u_2^4) + 
   u_1^6 (93 - 14 u_2^2 + u_2^4) \nonumber\\
 &+ 4 u_1^4 (36 - 13 u_2^2 + 4 u_2^4) - 
   3 u_1^3 (9 - 31 u_2^2 + 22 u_2^4) + 
   3 u_1^2 (27 - 42 u_2^2 + 31 u_2^4)) \,, \nonumber\\
\phi_{-2,4}(\bs{\tau}) =& \frac{\Psi_1(\tau_2)^4}{36 (1 + u_2)^3 L_1(\bs{u})L_2(\bs{u})\Psi_1(\tau_1)^2}(1 + u_1)^2 u_2^2 (9 - 9 u_2 - u_2^2 + u_2^3) (81 - 135 u_2^2 + 
         27 u_2^4\nonumber\\
         & - 5 u_2^6 + 2 u_1 (3 + u_2^2)^3 + 
         u_1^2 (45  -27 u_2^2 + 15 u_2^4 - 
            u_2^6)) \,, \nonumber\\
\phi_{-1,3}(\bs{\tau}) =&-\frac{(1 + u_1) ( u_2-3) ( u_2-1) u_2^2 (3 + u_2) \Psi_1(\tau_2)^3}{
 54 (1 + u_2)^2 (u_1^2 + u_2^2 - u_1 ( u_2^2-3)) \Psi_1(\tau_1)}\,, \nonumber\\
\phi^a_{0,2}(\bs{\tau}) =&-\frac{(u_2^2-3 ) (u_2^2+3 ) }{12 (1 + u_2)^2}\Psi_1(\tau_2)^2\,, \nonumber\\
\phi_{0,2}^b(\bs{\tau}) =& -\frac{
  u_2^2 (9 - 5 u_2^2 - u_1^2 (-5 + u_2^2) + 
     2 u_1 (3 + u_2^2)) \Psi_1(\tau_2)^2}{
  6 (1 + u_2)^2 (u_1^2 + u_2^2 - u_1 (-3 + u_2^2))}\,, \nonumber\\ 
\phi_{0,2}^c(\bs{\tau}) =&\frac{\Psi_1(\tau_2)^2}{36 (1 + u_2)^2 L_1(\bs{u})L_2(\bs{u})} (u_1^2 + u_2^2 - u_1 (u_2^2-3 ))(u_2^2 -3) ( u_2^2+3) (9 + 6 u_1 - 3 u_1^2 - 12 u_2\nonumber\\ 
&+ 
     24 u_1 u_2 + 4 u_1^2 u_2 + 3 u_2^2 + 2 u_1 u_2^2 - 
     u_1^2 u_2^2) (9 + 6 u_1 - 3 u_1^2 + 12 u_2 - 
     24 u_1 u_2 - 4 u_1^2 u_2 \nonumber\\ 
&+ 3 u_2^2 + 2 u_1 u_2^2 - 
     u_1^2 u_2^2) \,, \nonumber\\  
\phi_{0,4}(\bs{\tau}) =&  \frac{\Psi_1(\tau_2)^4}{216 (1 + u_2)^4 (u_1^2 + u_2^2 - u_1 (-3 + u_2^2))^2} u_2^2 (81 u_2^2 - 
   243 u_2^4 + 51 u_2^6 - 17 u_2^8 \nonumber\\
 &+ 
   u_1^4 (-153 + 51 u_2^2 - 27 u_2^4 + u_2^6) + 
   u_1^2 (-1215 + 612 u_2^2 - 218 u_2^4 + 68 u_2^6 - 15 u_2^8)\nonumber\\
 & + 
   u_1^3 (-837 + 342 u_2^2 - 28 u_2^4 + 10 u_2^6 + u_2^8) \nonumber\\
 &+ 
   u_1 (-243 - 270 u_2^2 + 84 u_2^4 - 114 u_2^6 + 
      31 u_2^8)) \,, \nonumber\\    
\phi_{1,1}(\bs{\tau}) =& -\frac{\Psi_1(\tau_2) \Psi_1(\tau_1)}{36 (1 + u_1) (1 + u_2) (u_1^2 + u_2^2 - u_1 (-3 + u_2^2))}(u_1^4 (3 + u_2^2) + 3 u_2^2 (3 + u_2^2)\nonumber\\
&+ 
       u_1^3 (27 - 20 u_2^2 + u_2^4) - 3 u_1 (9 - 20 u_2^2 + 3 u_2^4) + 
       u_1^2 (45 - 34 u_2^2 + 5 u_2^4)) ,\,, \nonumber\\
\phi_{2,2}(\bs{\tau}) = &-\frac{ \Psi_1(\tau_2)^2 \Psi_1(\tau_2)^2}{72 L_1(\bs{u})^2 L_2(\bs{u})^2 (1 + u_2)}(u_1-3)^2 u_1 (-3 + 2 u_1 + u_1^2) u_2^2\nonumber\\
& (9 - 9 u_2 - u_2^2 + 
   u_2^3) (u_1^8 (513 - 180 u_2^2 + 70 u_2^4 - 20 u_2^6 + u_2^8) + 
   8 u_1^7 (567 - 72 u_2^2\nonumber\\
& - 30 u_2^4 + 16 u_2^6 - u_2^8) + 
   216 u_1 (81 - 144 u_2^2 + 30 u_2^4 + 8 u_2^6 - 7 u_2^8) + 
   4 u_1^6 (5103 \nonumber\\
&- 1692 u_2^2 + 306 u_2^4 + 196 u_2^6 - 9 u_2^8) + 
   8 u_1^5 (4293 + 342 u_2^4 + 216 u_2^6 - 19 u_2^8) \nonumber\\
&+ 
   27 (243 - 540 u_2^2 + 210 u_2^4 - 60 u_2^6 + 19 u_2^8) + 
   24 u_1^3 (1539 - 1944 u_2^2\nonumber\\
& - 342 u_2^4 - 53 u_2^8) + 
   36 u_1^2 (-729 + 1764 u_2^2 + 306 u_2^4 - 188 u_2^6 + 63 u_2^8) \nonumber\\
&+ 
   2 u_1^4 (18387 - 4860 u_2^2 - 414 u_2^4 - 540 u_2^6 + 227 u_2^8)) \,, \nonumber\\
 \phi_{3,1}^a(\bs{\tau}) =&\Psi_1(\tau_1)^3 \Psi_1(\tau_2)\frac{(u_1-3 )^2 (u_1-1 ) u_1 (3 + u_1) (u_2^2-3) (3 + u_2^2) }{72 (1 + u_1) (1 + u_2) (3 u_1 + u_1^2 + u_2^2 -
    u_1 u_2^2)}\,, \nonumber\\
\phi_{3,1}^b(\bs{\tau}) =&\frac{\Psi_1(\tau_2) \Psi_1(\tau_1)^3  }{72 (1 + u_1) (1 + u_2) (u_1^2 + u_2^2 - 
   u_1 (-3 + u_2^2))^2}( u_1-3 )^2 u_1 (-3 + 2 u_1 + u_1^2)\nonumber\\
 & (u_2^2 (-27 + 10 u_2^2 + u_2^4) + 
   u_1^2 (-9 - 10 u_2^2 + 3 u_2^4) - 
   u_1 (27 + 3 u_2^2 + u_2^4 + u_2^6)) \,, \nonumber\\  
\phi_{4,2}(\bs{\tau}) =&  \frac{\Psi_1(\tau_2)^2 \Psi_1(\tau_1)^4}{864 (1 + u_1)^2 (1 + u_2)^2 (u_1^2 + u_2^2 - u_1 (-3 + u_2^2))^3} (-3 + 
   u_1)^2 (18 u_2^6 (-63 \nonumber\\
 &+ 30 u_2^2 + u_2^4) + 
   2 u_1^{10} (-9 - 30 u_2^2 + 7 u_2^4) - 
   27 u_1 u_2^4 (279 - 291 u_2^2 + 73 u_2^4 + 3 u_2^6) \nonumber\\
 &- 
   u_1^9 (243 + 657 u_2^2 - 291 u_2^4 + 31 u_2^6) + 
   2 u_1^8 (-702 - 1476 u_2^2 + 727 u_2^4 - 90 u_2^6 + 5 u_2^8) \nonumber\\
 &+ 
   6 u_1^2 u_2^2 (-1215 + 2430 u_2^2 - 2181 u_2^4 + 492 u_2^6 + 26 u_2^8) + 
   2 u_1^4 u_2^2 (-10854 + 6417 u_2^2 \nonumber\\
 &- 2455 u_2^4 + 611 u_2^6 + 
      41 u_2^8) + 
   2 u_1^6 (-3321 - 5499 u_2^2 + 2455 u_2^4 - 713 u_2^6 + 134 u_2^8) \nonumber\\
 &- 
   u_1^7 (4293 + 6759 u_2^2 - 3752 u_2^4 + 872 u_2^6 - 45 u_2^8 + u_2^{10}) - 
   3 u_1^5 (1215 + 7317 u_2^2 \nonumber\\
 &- 2526 u_2^4 - 842 u_2^6 + 271 u_2^8 + 
      5 u_2^{10})\nonumber\\
 & - 
   3 u_1^3 (729 - 3645 u_2^2 + 7848 u_2^4 - 3752 u_2^6 + 751 u_2^8 + 
      53 u_2^{10}))\,, \nonumber\\   
\phi_{1,3}^a(\bs{\tau}) =& -\Psi_1(\tau_1) \Psi_1(\tau_2)^3\frac{(3 + u_1^2) (u_1^2 + 6 u_1 -3) ( u_2-3) (u_2-1 ) u_2^2 (3 + u_2) }{
 216 (1 + u_1) (1 + u_2)^2 (3 u_1 + u_1^2 + u_2^2 - u_1 u_2^2)}\,, \nonumber\\  
\phi_{1,3}^b(\bs{\tau}) =&-\frac{ \Psi_1(\tau_2)^3 \Psi_1(\tau_1)}{
 108 (1 + u_1) (1 + u_2)^2 (u_1^2 + u_2^2 - u_1 (-3 + u_2^2))^2}u_2^2 (9 - 9 u_2 - u_2^2 + u_2^3) \nonumber\\
 &(4 u_1^5 + u_1^6 - 9 u_2^2 + 
    12 u_1 u_2^2 + u_1^4 (2 - 5 u_2^2) + u_1^2 (45 - 2 u_2^2) + 
    4 u_1^3 (3 + u_2^2))\,, \nonumber\\      
\phi_{2,4}(\bs{\tau}) =&\frac{-\Psi_1(\tau_2)^4 \Psi_1(\tau_1)^2}{2592 (1 + u_1)^2 (1 + u_2)^4 (u_1^2 + u_2^2 - 
     u_1 (-3 + u_2^2))^3} u_2^2 (u_1^{10} (531 - 33 u_2^2 + 9 u_2^4 + 
      5 u_2^6)\nonumber\\
 & - 9 u_2^4 (405 + 81 u_2^2 - 33 u_2^4 + 59 u_2^6) - 
   2 u_1^9 (-2133 + 531 u_2^2 + 129 u_2^4 - 71 u_2^6 + 8 u_2^8)\nonumber\\
 & + 
   18 u_1 u_2^2 (-1944 + 1917 u_2^2 - 387 u_2^4 - 177 u_2^6 + 79 u_2^8) \nonumber\\
 &+ 
   u_1^6 (25920 - 3960 u_2^2 - 2138 u_2^4 + 1454 u_2^6 + 594 u_2^8 - 
      366 u_2^{10}) \nonumber\\
 &+ 
   u_1^4 (266814 - 48114 u_2^2 - 13086 u_2^4 + 2138 u_2^6 + 440 u_2^8 - 
      320 u_2^{10})\nonumber\\
 & + 
   u_1^8 (7506 - 7902 u_2^2 + 2683 u_2^4 - 737 u_2^6 - 17 u_2^8 + 
      3 u_2^{10}) \nonumber\\
 &+ 
   2 u_1^7 (-3483 - 5355 u_2^2 + 5990 u_2^4 - 1554 u_2^6 + 293 u_2^8 + 
      13 u_2^{10})\nonumber\\
 & - 
   6 u_1^3 (-9477 - 23733 u_2^2 + 13986 u_2^4 - 5990 u_2^6 + 595 u_2^8 + 
      43 u_2^{10}) \nonumber\\
 &+ 
   6 u_1^5 (34749 - 13635 u_2^2 + 564 u_2^4 + 188 u_2^6 - 505 u_2^8 + 
      143 u_2^{10}) \nonumber\\
 &- 
   3 u_1^2 (6561 - 4131 u_2^2 - 19899 u_2^4 + 8049 u_2^6 - 2634 u_2^8 + 
      278 u_2^{10})) \,.
\end{align}

% !TEX root = main.tex

\section{Independence of the differential forms}
\label{app:independence}

In this appendix we present the details of the proof that the differential forms $\omega_{ij} = \rd\tau_1\,\omega_{ij}^1+\rd\tau_2\,\omega_{ij}^2$ defined in appendix~\ref{app:diff_forms_can_eq} define independent cohomology classes. We start by discussing some technical result about connections and gauge transformations before we turn to the proof of the independence of the differential forms.

\subsection{Connections and gauge transformations}
\label{app:independence_1}
We consider a differential equation of the form
\beq
\rd \bI(x)  = \bOmega(x)\bI(x)\,,
\eeq
 where $x$ is a complex variable and $\bI(x)$ is a $n$-dimensional vector. The differential equation matrix has the form
\beq
\bOmega(x) = \sum_{p=1}^N\bA_p\,\omega_p\,,
\eeq
where $A_p$ are some non-commutative variables (e.g., matrices). The differential forms are $\omega_p = \rd x\,\xi_{p}(x)$, where the $ \xi_{p}$ are taken from some function field $\cF$ that is differentially closed (meaning that it is closed under taking derivatives in $x$) and whose field of constants is $\mathbb{C}$. We assume that the $\omega_p$ define independent cohomology classes \emph{with respect to $\cF$}, i.e., we assume that whenever there are $c_1,\ldots,c_N\in\mathbb{C}$ and $f\in \cF$ such that
\beq
\sum_{p=1}^Nc_p\,\omega_p = \rd f\,,
\eeq 
then necessarily $c_1=\ldots=c_N=0$ and $f$ is a constant. 
The solution to this equation can be written in terms of the path-ordered exponential in eqs.~\eqref{eq:Pexp_def} and~\eqref{eq:Pexp_def_exp}, and the fact that the differential forms define independent cohomology classes implies that the iterated integrals in eq.~\eqref{eq:Pexp_def_exp} formed from of the letters $\omega_1,\ldots,\omega_N$ are linearly independent over $\cF$~\cite{deneufchatel:hal-00558773}.

Consider a path $\gamma$ from a point $x_0$ to a generic point $x$ (and which avoids possible singularities from any of the $\xi_p$), and define
\beq
\mathbb{I}(x) = \int_{\gamma} \omega_N = \int_{x_0}^x\rd x'\,\xi_N(x')\,.
\eeq
Due to our assumption that the $\omega_p$ define independent cohomology classes, it follows that $\mathbb{I}$ does not lie in $\cF$. 
We can then consider the field $\cG := \cF(\mathbb{I})$ whose elements are rational functions in $\mathbb{I}$ with coefficients in $\cF$. Note that since $\partial_x\mathbb{I} = \xi_N\in\cF$, $\cG$ is also differentially closed, and its field of constants is still $\mathbb{C}$.

Consider the gauge transformation
\beq\label{eq:J_gauge}
\bJ(x) = e^{-A_N\mathbb{I}(x)}\bI(x)\,.
\eeq
In this gauge, the differential equation takes the form 
\beq
\rd\bJ(c) = e^{-A_N\mathbb{I}(x)}\widetilde{\bOmega}(x)e^{A_N\mathbb{I}(x)}\bJ(x)\,,\qquad \widetilde{\bOmega}(x) = \sum_{p=1}^{N-1}\bA_p\,\omega_p\,,
\eeq
and the differential equation matrix is given by
\beq
e^{-A_N\mathbb{I}(x)}\widetilde{\bOmega}e^{A_N\mathbb{I}(x)} = \sum_{r=0}^\infty\sum_{p=1}^{N-1}\frac{(-1)^r}{r!}\operatorname{ad}_{A_N}^r(A_p)\,\omega_p^{(r)}\,,
\eeq
with $\operatorname{ad}_{A_N}(A_p)=[A_N,A_p]$ and 
\beq
\omega_p^{(r)} := \mathbb{I}(x)^r\omega_p\,.
\eeq
Our goal is to show that the differential forms $\omega_p^{(r)}$, $r\ge 0$ and $1\le p<N$, which are the letters that appear in the iterated integrals in the path-ordered exponential that solves eq.~\eqref{eq:J_gauge}, define independent cohomology classes with respect to $\cG$, i.e., whenever there are $c_{rp}\in\mathbb{C}$ and $g\in\cG$ such that
\beq\label{eq:gauge_1}
\sum_{r\ge 0}\sum_{p=1}^{N-1}c_{rp}\,\omega_p^{(r)} = \rd g\,,
\eeq
then necessarily $c_{rp}=0$ and $g$ is a constant.
To see that this is true, we integrate eq.~\eqref{eq:gauge_1} over the path $\gamma$ used to define $\mathbb{I}$. We obtain
\beq\label{eq:gauge_2}
\sum_{r\ge 0}\sum_{p=1}^{N-1}c_{rp}\,I_{\gamma}\big(\omega_p^{(r)}\big) = g(x) - g(x_0)=:\tilde{g}(x)\,,
\eeq
and $I_{\gamma}(\omega)$ denotes the iterated integral of length one, cf.~eq.~\eqref{eq:iterated_int_def}. Using the fact that iterated integrals form a shuffle algebra, we find 
\beq\bsp
I_{\gamma}\big(\omega_p^{(r)}\big) &\,= \int_{\gamma}\omega_p^{(r)} = \int_{\gamma}\omega_p\,\mathbb{I}(x)^r=\int_{\gamma}\omega_p\left(\int_{\gamma}\omega_N\right)^r=r!\int_{\gamma}\omega_p\underbrace{\omega_N\cdots\omega_N}_{r \textrm{ times}}\\
&\,=r!\, I_{\gamma}(\omega_p,\underbrace{\omega_N,\ldots,\omega_N}_{r\textrm{ times}}) = r!\, I_{\gamma}(\omega_p,\omega_N^r)\,,
\esp\eeq
where in the last equality we introduced the simplified notation
\beq
I_{\gamma}(\omega_p,\omega_N^r) := I_{\gamma}(\omega_p,\underbrace{\omega_N,\ldots,\omega_N}_{r\textrm{ times}})\,.
\eeq
Since $g\in\cG$, also $\tilde{g}\in\cG$, and so we can write
\beq
\tilde{g} = \frac{P(\mathbb{I})}{Q(\mathbb{I})}\,,
\eeq
where $P$ and $Q$ are polynomials,
\beq\bsp
P(\mathbb{I}) &\,= \sum_{k=0}^{d_P}\frac{p_k}{k!}\,\mathbb{I}^k = \sum_{k=0}^{d_P}p_k\,I_{\gamma}(\omega_N^k) \,,\\
 Q(\mathbb{I}) &\,= \sum_{k=0}^{d_Q}\frac{q_k}{k!}\,\mathbb{I}^k = \sum_{k=0}^{d_Q}q_k\,I_{\gamma}(\omega_N^k)\,,
\esp\eeq
with $p_k,q_k\in \cF$. Equation~\eqref{eq:gauge_2} can then be cast in the form
\beq\label{eq:gauge_3}
\sum_{r\ge 0}\sum_{p=1}^{N-1}\sum_{k=0}^{d_Q}\,r!\,c_{rp}\,q_k\,I_{\gamma}(\omega_N^k)I_{\gamma}(\omega_p,\omega_N^r) =  \sum_{k=0}^{d_P}p_k\,I_{\gamma}(\omega_N^k)\,.
\eeq
Using the shuffle algebra structure of iterated integrals, we can write $I_{\gamma}(\omega_N^k)I_{\gamma}(\omega_p,\omega_N^r)$ again as a linear combination (with integer coefficients) of iterated integrals of length $k+r+1>0$. We then see that both sides of eq.~\eqref{eq:gauge_3} are linear combinations with coefficients in $\cF$ of iterated integrals made from the letters $\omega_1,\ldots,\omega_N$.Every iterated integral on the left-hand side has length at least 1 and involves exactly one letter $\omega_p$ with $p\neq N$, while those on the right-hand side only involve the letter $\omega_N$, except for the integral with $k=0$, which has length 0. Since we know that these iterated integrals are linearly independent over $\cF$, both sides must vanish independently. In particular this means $p_0=\ldots=p_{d_P}=0$, and so $P=0$. This implies $\tilde{g}=0$, or equivalently $g$ must be a constant. Equation~\eqref{eq:gauge_2} then takes the form
\beq
\sum_{r\ge 0}\sum_{p=1}^{N-1}c_{rp}\,I_{\gamma}\big(\omega_p^{(r)}\big)  = \sum_{r\ge 0}\sum_{p=1}^{N-1}r!\,c_{rp}\,I_{\gamma}(\omega_p,\omega_N^r) =0\,.
\eeq
The linear independence of the $I_{\gamma}(\omega_p,\omega_N^r)$ then implies $c_{rp} = 0$, which finishes the proof.

%%%%%%%%%%%%%%%%%%%%%%%%%%%%%%%

\subsection{Independence of the differential forms}
We now turn to the proof that the differential forms $\omega_{ij}$ that appear in the canonical differential equation in section~\ref{sec:canonicaldeq} define independent cohomology classes.

We start by assuming that $\tau_2$ has been fixed to some constant generic value, and we only consider the one-forms $\rd\tau_1\,\omega_{ij}^1$. We know from appendix~\ref{app:diff_forms_can_eq} that the $\omega_{ij}^1$ are constructed from the $\eps$-functions $\mathbb{I}_k(\cdot,\tau_2)$ 
in eq.~\eqref{new_int} and the meromorphic modular forms $\phi_{k,l}(\cdot,\tau_2),\phi_{k,l}^r(\cdot,\tau_2)\in\cM_k(\Gamma^1(12))$  from appendix~\ref{app:diff_forms_can_eq}. From the explicit expression in appendix~\ref{app:diff_forms_can_eq}, we see that all relevant meromorphic modular forms for $\Gamma^1(12)$ have positive weight (seen as a modular form in $\tau_1$), and we can group them into three classes:
\begin{itemize}
\item \textbf{Class I:} Functions of class I have at most simple poles:
\beq
1\,,\phi_{3,-1}\,,\phi_{4,-2}\,,\phi_{2,0}^a\,,\phi_{3,1}^a\,,\phi_{0,2}^b\,,\phi_{2,0}^b\,,\phi_{0,2}^c\,,\phi_{2,0}^c\,.
\eeq
\item \textbf{Class II:} Functions of class II have at least double poles double or at most triple poles:
\beq
\phi_{2,2}\,,\phi_{4,0}\,,\phi_{4,2}\,,\phi_{3,1}^b\,.
\eeq
\item \textbf{Class III:} Functions of class III do not enter the $\omega_{ij}^1$ directly, but only through the $\eps$-functions in eqs.~\eqref{new_int} and~\eqref{new_int}:\footnote{We of course take the derivative in $\tau_2$ before fixing $\tau_2$ to a constant value.}
\beq
\partial_{\tau_2}\phi_{3,-1}(\cdot,\tau_2)\,,\qquad \partial_{\tau_2}\phi_{4,-2}(\cdot,\tau_2)\,,\qquad \partial_{\tau_2}^2\phi_{4,-2}(\cdot,\tau_2)\,.
\eeq
We know from the discussion in section~\ref{sec:canonicaldeq} that these functions have double or triple poles, respectively.
\end{itemize}
By inspection, we see that the pole orders of these meromorphic modular forms satisfy the criterion to lie in the vector space $\widetilde{\cM}(\Gamma^1(12))$ defined in appendix~\ref{app:modular}. We denote by $\widetilde{\cN}_1\subseteq\widetilde{\cM}(\Gamma^1(12))$ the linear span of these meromorphic modular forms. From eq.~\eqref{eq:proof_eq_2} we see that
\beq\label{eq:N1_cap}
\widetilde{\cN}_1\cap \partial_{\tau_1}\mathcal{QM}(\Gamma^1(12)) = \{0\}\,.
\eeq
Said differently, if we pick a basis $\{f_p\}$ for $\widetilde{\cN}_1$, then the differential forms $\rd\tau_1\,f_p(\tau_1)$ define independent cohomology classes (seen as functions of $\tau_1$ only).

So far, we have only considered meromorphic modular forms. However, the functions of class III do not appear directly in the definition of the $\omega_{ij}^1$, but they only enter through the $\eps$-functions $\mathbb{I}_k(\cdot,\tau_2)$. Applying the result from appendix~\ref{app:independence_1} to the functions of class III one at the time, we can eliminate them one after and replace them with the $\eps$-functions. We consider the vector space $\widetilde{\cN}_1^{\eps}$ generated by products of $\eps$-functions with a function of class I and II.\footnote{Note that along the way we also introduce products of functions of class III and $\eps$-functions. Those, however, do not appear in the differential forms from section~\ref{app:diff_forms_can_eq}.} Equation~\eqref{eq:N1_cap} together the analysis from appendix~\ref{app:independence_1} then implies that 
\beq\label{eq:N1e_cap}
\widetilde{\cN}_1^{\eps}\cap \partial_{\tau_1}\mathcal{QM}(\Gamma^1(12))(\mathbb{I}_1,\mathbb{I}_2,\mathbb{I}_3)= \{0\}\,,
\eeq
where the elements of $\mathcal{QM}(\Gamma^1(12))(\mathbb{I}_1,\mathbb{I}_2,\mathbb{I}_3)$ are rational functions in $\eps$-functions with coefficients that are quasi-modular forms.

Let $\widetilde{\cN}_1^{\omega}$ denote the linear span of the functions $\omega_{ij}^1$. By construction, we have $\widetilde{\cN}_1^{\omega}\subseteq\widetilde{\cN}_1^{\eps}$, and so eq.~\eqref{eq:N1e_cap} implies
\beq\label{eq:N1o_cap}
\widetilde{\cN}_1^{\omega}\cap \partial_{\tau_1}\mathcal{QM}(\Gamma^1(12))(\mathbb{I}_1,\mathbb{I}_2,\mathbb{I}_3)= \{0\}\,.
\eeq
We can of course repeat exactly the same construction starting from the $\omega_{ij}^2$. If we denote their span by $\widetilde{\cN}_2^{\omega}$, we arrive at the conclusion that
\beq\label{eq:N2o_cap}
\widetilde{\cN}_2^{\omega}\cap \partial_{\tau_2}\mathcal{QM}(\Gamma_1(12))(\mathbb{I}_1,\mathbb{I}_2,\mathbb{I}_3)= \{0\}\,.
\eeq

We are now in the position to prove that the differential forms $\omega_{ij}$ define independent cohomology classes. Assume that there are constants $c_{ij}$ and a function 
\beq
F\in \mathcal{QM}(\Gamma^1(12),\Gamma_1(12))(\mathbb{I}_1,\mathbb{I}_2,\mathbb{I}_3)
\eeq
 such that
\beq\label{eq:indep_1}
\sum_{ij}c_{ij}\,\omega_{ij} = \rd F(\tau_1,\tau_2)\,,
\eeq
where the sum runs over a linearly independent set of the $\omega_{ij}$.\footnote{Such a set is easy to obtain, by simply resolving trivial linear relations among the $\omega_{ij}$.} Equation~\eqref{eq:indep_1} is equivalent to the system
\beq
\sum_{ij}c_{ij}\,\omega_{ij}^k = \partial_{\tau_k}F(\tau_1,\tau_2)\,, \qquad k=1,2\,.
\eeq
We focus on $k=1$, and we fix $\tau_2$ to a constant generic value. The left-hand side lies in $\widetilde{\cN}_1^{\omega}$ by construction, while the right-hand side lies in $ \partial_{\tau_1}\mathcal{QM}(\Gamma^1(12))(\mathbb{I}_1,\mathbb{I}_2,\mathbb{I}_3)$. Equation~\eqref{eq:N1o_cap} then implies $\partial_{\tau_1}F=0$. Applying exactly the same reasoning to $k=2$, we conclude that also $\partial_{\tau_2}F=0$, and so $\rd F=0$. Hence, eq.~\eqref{eq:indep_1} reduces to
\beq
\sum_{ij}c_{ij}\,\omega_{ij} = 0\,,
\eeq
and so $c_{ij}=0$ from the linear independence of the $\omega_{ij}$.

\end{appendix}

\bibliographystyle{jhep}
\bibliography{bananabib}

@article{magnetic4,
	author = {Allan, Michael and Long, Ling and Saad, Hasan},
	eprint = {2511.05718},
	month = {11},
	primaryclass = {math.NT},
	title = {{Atkin and Swinnerton-Dyer congruences for meromorphic modular forms}},
	archiveprefix = {arXiv},
	year = 2025}

@article{Baune:2024ber,
	archiveprefix = {arXiv},
	author = {Baune, Konstantin and Broedel, Johannes and Im, Egor and Lisitsyn, Artyom and Moeckli, Yannis},
	doi = {10.21468/SciPostPhys.18.3.093},
	eprint = {2409.08208},
	journal = {SciPost Phys.},
	month = {9},
	pages = {093},
	primaryclass = {hep-th},
	title = {{Higher-genus Fay-like identities from meromorphic generating functions}},
	volume = {18},
	year = {2025}}

@article{EnriquezZerbini,
	archiveprefix = {arXiv},
	author = {Enriquez, Benjamin and Zerbini, Federico},
	doi = {10.4153/S0008414X24001068},
	eprint = {2307.01833},
	journal = {Canadian Journal of Mathematics},
	pages = {1--36},
	primaryclass = {math.AG},
	title = {Elliptic hyperlogarithms},
	year = {2025}}

@article{Bree:2025tug,
	archiveprefix = {arXiv},
	author = {Bree, Iris and others},
	eprint = {2511.15381},
	month = {11},
	primaryclass = {hep-th},
	title = {{New algorithms for Feynman integral reduction and $\varepsilon$-factorised differential equations}},
	year = {2025}}

@article{Weinzierl:2020fyx,
	archiveprefix = {arXiv},
	author = {Weinzierl, Stefan},
	doi = {10.1016/j.nuclphysb.2021.115309},
	eprint = {2011.07311},
	journal = {Nucl. Phys. B},
	pages = {115309},
	primaryclass = {hep-th},
	title = {{Modular transformations of elliptic Feynman integrals}},
	volume = {964},
	year = {2021}}

@incollection{deneufchatel:hal-00558773,
	author = {Deneufch{\^a}tel, Matthieu and Duchamp, G{\'e}rard H.E. and Hoang Ngoc Minh, Vincel and Solomon, Allan I.},
	booktitle = {{Lecture Notes in Computer Science, Volume 6742 (2011),}},
	hal_id = {hal-00558773},
	hal_version = {v4},
	month = May,
	title = {{Independence of hyperlogarithms over function fields via algebraic combinatorics.}},
	url = {https://hal.science/hal-00558773},
	year = {2011}}

@article{Maggio:2025jel,
	archiveprefix = {arXiv},
	author = {Maggio, Sara and Sohnle, Yoann},
	doi = {10.1007/JHEP10(2025)202},
	eprint = {2504.17757},
	journal = {JHEP},
	pages = {202},
	primaryclass = {hep-th},
	reportnumber = {BONN-TH-2025-16, UUITP--14/25},
	title = {{On canonical differential equations for Calabi-Yau multi-scale Feynman integrals}},
	volume = {10},
	year = {2025}}

@article{Duhr:2025lbz,
	archiveprefix = {arXiv},
	author = {Duhr, Claude and Maggio, Sara and Nega, Christoph and Sauer, Benjamin and Tancredi, Lorenzo and Wagner, Fabian J.},
	doi = {10.1007/JHEP06(2025)128},
	eprint = {2503.20655},
	journal = {JHEP},
	pages = {128},
	primaryclass = {hep-th},
	reportnumber = {BONN-TH-2025-11, TUM-HEP 1559/25, HU-EP-25/13-RTG},
	title = {{Aspects of canonical differential equations for Calabi-Yau geometries and beyond}},
	volume = {06},
	year = {2025}}

@article{Mishnyakov:2024rmb,
	archiveprefix = {arXiv},
	author = {Mishnyakov, V. and Morozov, A. and Reva, M. and Suprun, P.},
	doi = {10.1142/S0217751X25500174},
	eprint = {2404.03069},
	journal = {Int. J. Mod. Phys. A},
	number = {07},
	pages = {2550017},
	primaryclass = {hep-th},
	title = {{From equations in coordinate space to Picard{\textendash}Fuchs and back}},
	volume = {40},
	year = {2025}}

@article{DHoker:2023vax,
	archiveprefix = {arXiv},
	author = {D'Hoker, Eric and Hidding, Martijn and Schlotterer, Oliver},
	doi = {10.4310/cntp.250531031558},
	eprint = {2306.08644},
	journal = {Commun. Num. Theor. Phys.},
	number = {2},
	pages = {355--413},
	primaryclass = {hep-th},
	title = {{Constructing polylogarithms on higher-genus Riemann surfaces}},
	volume = {19},
	year = {2025}}

@article{Brown:2025zsw,
	archiveprefix = {arXiv},
	author = {Brown, Francis and Fonseca, Tiago J.},
	eprint = {2508.04844},
	month = {8},
	primaryclass = {math.NT},
	title = {{Single-valued periods of meromorphic modular forms and a motivic interpretation of the Gross-Zagier conjecture}},
	year = {2025}}

@article{Pogel:2024sdi,
	archiveprefix = {arXiv},
	author = {P{\"o}gel, Sebastian and Wang, Xing and Weinzierl, Stefan and Wu, Konglong and Xu, Xiaofeng},
	doi = {10.1007/JHEP09(2024)084},
	eprint = {2407.08799},
	journal = {JHEP},
	pages = {084},
	primaryclass = {hep-th},
	reportnumber = {MITP/24-059, TUM-HEP-1513/24},
	title = {{Self-dualities and Galois symmetries in Feynman integrals}},
	volume = {09},
	year = {2024}}

@article{Chaubey:2025adn,
	archiveprefix = {arXiv},
	author = {Chaubey, Ekta and Sotnikov, Vasily},
	doi = {10.1103/4fjc-lfnx},
	eprint = {2504.20897},
	journal = {Phys. Rev. Lett.},
	number = {10},
	pages = {101903},
	primaryclass = {hep-th},
	reportnumber = {BONN-TH-2025-18, ZU-TH 29/25},
	title = {{Elliptic Leading Singularities and Canonical Integrands}},
	volume = {135},
	year = {2025}}

@article{Chen:2025hzq,
	archiveprefix = {arXiv},
	author = {Chen, Jiaqi and Yang, Li Lin and Zhang, Yiyang},
	eprint = {2503.23720},
	month = {3},
	primaryclass = {hep-th},
	title = {{On an approach to canonicalizing elliptic Feynman integrals}},
	year = {2025}}

@article{e-collaboration:2025frv,
	archiveprefix = {arXiv},
	author = {Bree, Iris and others},
	collaboration = {{\ensuremath{\varepsilon}}-collaboration},
	eprint = {2506.09124},
	month = {6},
	primaryclass = {hep-th},
	title = {{The geometric bookkeeping guide to Feynman integral reduction and $\varepsilon$-factorised differential equations}},
	year = {2025}}

@article{Bogner:2019lfa,
	archiveprefix = {arXiv},
	author = {Bogner, Christian and M{\"u}ller-Stach, Stefan and Weinzierl, Stefan},
	doi = {10.1016/j.nuclphysb.2020.114991},
	eprint = {1907.01251},
	journal = {Nucl. Phys. B},
	pages = {114991},
	primaryclass = {hep-th},
	title = {{The unequal mass sunrise integral expressed through iterated integrals on $\overline{\mathcal M}_{1,3}$}},
	volume = {954},
	year = {2020}}

@article{Adams:2015ydq,
	archiveprefix = {arXiv},
	author = {Adams, Luise and Bogner, Christian and Weinzierl, Stefan},
	doi = {10.1063/1.4944722},
	eprint = {1512.05630},
	journal = {J. Math. Phys.},
	number = {3},
	pages = {032304},
	primaryclass = {hep-ph},
	reportnumber = {MITP-15-114, MaPhy-AvH-2015-21},
	title = {{The iterated structure of the all-order result for the two-loop sunrise integral}},
	volume = {57},
	year = {2016}}

@article{Adams:2014vja,
	archiveprefix = {arXiv},
	author = {Adams, Luise and Bogner, Christian and Weinzierl, Stefan},
	doi = {10.1063/1.4896563},
	eprint = {1405.5640},
	journal = {J. Math. Phys.},
	number = {10},
	pages = {102301},
	primaryclass = {hep-ph},
	title = {{The two-loop sunrise graph in two space-time dimensions with arbitrary masses in terms of elliptic dilogarithms}},
	volume = {55},
	year = {2014}}

@article{Adams:2013nia,
	archiveprefix = {arXiv},
	author = {Adams, Luise and Bogner, Christian and Weinzierl, Stefan},
	doi = {10.1063/1.4804996},
	eprint = {1302.7004},
	journal = {J. Math. Phys.},
	pages = {052303},
	primaryclass = {hep-ph},
	title = {{The two-loop sunrise graph with arbitrary masses}},
	volume = {54},
	year = {2013}}

@article{Bloch:2013tra,
	archiveprefix = {arXiv},
	author = {Bloch, Spencer and Vanhove, Pierre},
	doi = {10.1016/j.jnt.2014.09.032},
	eprint = {1309.5865},
	journal = {J. Number Theor.},
	pages = {328--364},
	primaryclass = {hep-th},
	title = {{The elliptic dilogarithm for the sunset graph}},
	volume = {148},
	year = {2015}}

@article{Broedel:2017siw,
	archiveprefix = {arXiv},
	author = {Broedel, Johannes and Duhr, Claude and Dulat, Falko and Tancredi, Lorenzo},
	doi = {10.1103/PhysRevD.97.116009},
	eprint = {1712.07095},
	journal = {Phys. Rev. D},
	number = {11},
	pages = {116009},
	primaryclass = {hep-ph},
	reportnumber = {CERN-TH-2017-274, CP3-17-58, HU-EP-17-30, HU-Mathematik-2017-10, SLAC-PUB-17195},
	title = {{Elliptic polylogarithms and iterated integrals on elliptic curves II: an application to the sunrise integral}},
	volume = {97},
	year = {2018}}

@article{Brown:2014pnb,
	archiveprefix = {arXiv},
	author = {Brown, Francis},
	eprint = {1407.5167},
	month = {7},
	primaryclass = {math.NT},
	title = {{Multiple Modular Values and the relative completion of the fundamental group of $M_{1,1}$}},
	year = {2014}}

@article{Broedel:2017kkb,
	archiveprefix = {arXiv},
	author = {Broedel, Johannes and Duhr, Claude and Dulat, Falko and Tancredi, Lorenzo},
	doi = {10.1007/JHEP05(2018)093},
	eprint = {1712.07089},
	journal = {JHEP},
	pages = {093},
	primaryclass = {hep-th},
	reportnumber = {CERN-TH-2017-273, CP3-17-57, HU-EP-17-29, HU-Mathematik-2017-09, SLAC-PUB-17194},
	title = {{Elliptic polylogarithms and iterated integrals on elliptic curves. Part I: general formalism}},
	volume = {05},
	year = {2018}}

@article{Ablinger:2011te,
	archiveprefix = {arXiv},
	author = {Ablinger, Jakob and Blumlein, Johannes and Schneider, Carsten},
	doi = {10.1063/1.3629472},
	eprint = {1105.6063},
	journal = {J. Math. Phys.},
	pages = {102301},
	primaryclass = {math-ph},
	reportnumber = {DESY-11-033, DO-TH-11-12, SFB-CPP-11-24, LPN-11-24, DESY-11--033, DO--TH-11--12},
	title = {{Harmonic Sums and Polylogarithms Generated by Cyclotomic Polynomials}},
	volume = {52},
	year = {2011}}

@article{Gehrmann:2000zt,
	archiveprefix = {arXiv},
	author = {Gehrmann, T. and Remiddi, E.},
	doi = {10.1016/S0550-3213(01)00057-8},
	eprint = {hep-ph/0008287},
	journal = {Nucl. Phys. B},
	pages = {248--286},
	reportnumber = {KA-TTP-00-20},
	title = {{Two loop master integrals for gamma* ---{\ensuremath{>}} 3 jets: The Planar topologies}},
	volume = {601},
	year = {2001}}

@article{Duhr:2025xyy,
	archiveprefix = {arXiv},
	author = {Duhr, Claude and Maggio, Sara and Porkert, Franziska and Semper, Cathrin and Sohnle, Yoann and Stawinski, Sven F.},
	eprint = {2509.17787},
	month = {9},
	primaryclass = {hep-th},
	reportnumber = {BONN-TH/2025-30, UUITP--27/25},
	title = {{Canonical differential equations and intersection matrices}},
	year = {2025}}

@article{Matthes2022IteratedPrimitives,
	author = {Matthes, Nils},
	doi = {10.1090/tran/8538},
	journal = {Transactions of the American Mathematical Society},
	note = {arXiv:2101.11491},
	number = {2},
	pages = {1443--1460},
	title = {Iterated primitives of meromorphic quasimodular forms for ${\mathrm{SL}}_{2}(\mathbb{Z})$},
	volume = {375},
	year = {2022}}

@article{Joyce1973SimpleCubicLGF,
	author = {Joyce, G. S.},
	doi = {10.1098/rsta.1973.0018},
	journal = {Philosophical Transactions of the Royal Society of London. Series A, Mathematical and Physical Sciences},
	number = {1235},
	pages = {583--610},
	title = {On the simple cubic lattice Green function},
	volume = {273},
	year = {1973}}

@article{Frellesvig:2017aai,
	archiveprefix = {arXiv},
	author = {Frellesvig, Hjalte and Papadopoulos, Costas G.},
	doi = {10.1007/JHEP04(2017)083},
	eprint = {1701.07356},
	journal = {JHEP},
	pages = {083},
	primaryclass = {hep-ph},
	title = {{Cuts of Feynman Integrals in Baikov representation}},
	volume = {04},
	year = {2017}}

@article{Broedel:2018qkq,
	archiveprefix = {arXiv},
	author = {Broedel, Johannes and Duhr, Claude and Dulat, Falko and Penante, Brenda and Tancredi, Lorenzo},
	doi = {10.1007/JHEP01(2019)023},
	eprint = {1809.10698},
	journal = {JHEP},
	pages = {023},
	primaryclass = {hep-th},
	reportnumber = {CP3-18-58, CERN-TH-2018-211, HU-Mathematik-2018-09, HU-EP-18/29, SLAC-PUB-17336},
	title = {{Elliptic Feynman integrals and pure functions}},
	volume = {01},
	year = {2019}}

@article{Arkani-Hamed:2010pyv,
	archiveprefix = {arXiv},
	author = {Arkani-Hamed, Nima and Bourjaily, Jacob L. and Cachazo, Freddy and Trnka, Jaroslav},
	doi = {10.1007/JHEP06(2012)125},
	eprint = {1012.6032},
	journal = {JHEP},
	pages = {125},
	primaryclass = {hep-th},
	title = {{Local Integrals for Planar Scattering Amplitudes}},
	volume = {06},
	year = {2012}}

@article{Kotikov:2010gf,
	archiveprefix = {arXiv},
	author = {Kotikov, A. V.},
	eprint = {1005.5029},
	month = {5},
	pages = {150--174},
	primaryclass = {hep-th},
	title = {{The Property of maximal transcendentality in the N=4 Supersymmetric Yang-Mills}},
	year = {2010}}

@article{Kotikov:1990kg,
	author = {Kotikov, A. V.},
	doi = {10.1016/0370-2693(91)90413-K},
	journal = {Phys. Lett. B},
	pages = {158--164},
	reportnumber = {ITF-90-31E},
	title = {{Differential equations method: New technique for massive Feynman diagrams calculation}},
	volume = {254},
	year = {1991}}

@article{Kotikov:1991hm,
	author = {Kotikov, A. V.},
	doi = {10.1016/0370-2693(91)90834-D},
	journal = {Phys. Lett. B},
	pages = {314--322},
	title = {{Differential equations method: The Calculation of vertex type Feynman diagrams}},
	volume = {259},
	year = {1991}}

@article{Kotikov:1991pm,
	author = {Kotikov, A. V.},
	doi = {10.1016/0370-2693(91)90536-Y},
	journal = {Phys. Lett. B},
	note = {[Erratum: Phys.Lett.B 295, 409--409 (1992)]},
	pages = {123--127},
	title = {{Differential equation method: The Calculation of N point Feynman diagrams}},
	volume = {267},
	year = {1991}}

@article{Chetyrkin:1981qh,
	author = {Chetyrkin, K. G. and Tkachov, F. V.},
	doi = {10.1016/0550-3213(81)90199-1},
	journal = {Nucl. Phys. B},
	pages = {159--204},
	title = {{Integration by parts: The algorithm to calculate $\beta$-functions in 4 loops}},
	volume = {192},
	year = {1981}}

@article{Tkachov:1981wb,
	author = {Tkachov, F. V.},
	doi = {10.1016/0370-2693(81)90288-4},
	journal = {Phys. Lett. B},
	pages = {65--68},
	title = {{A theorem on analytical calculability of 4-loop renormalization group functions}},
	volume = {100},
	year = {1981}}

@article{Bollini:1972ui,
	author = {Bollini, C. G. and Giambiagi, J. J.},
	doi = {10.1007/BF02895558},
	journal = {Nuovo Cim. B},
	pages = {20--26},
	title = {{Dimensional Renormalization: The Number of Dimensions as a Regularizing Parameter}},
	volume = {12},
	year = {1972}}

@article{tHooft:1972tcz,
	author = {'t Hooft, Gerard and Veltman, M. J. G.},
	doi = {10.1016/0550-3213(72)90279-9},
	journal = {Nucl. Phys. B},
	pages = {189--213},
	title = {{Regularization and Renormalization of Gauge Fields}},
	volume = {44},
	year = {1972}}

@article{Cicuta:1972jf,
	author = {Cicuta, G. M. and Montaldi, E.},
	doi = {10.1007/BF02756527},
	journal = {Lett. Nuovo Cim.},
	pages = {329--332},
	title = {{Analytic renormalization via continuous space dimension}},
	volume = {4},
	year = {1972}}

@article{Henn:2013pwa,
	archiveprefix = {arXiv},
	author = {Henn, Johannes M.},
	doi = {10.1103/PhysRevLett.110.251601},
	eprint = {1304.1806},
	journal = {Phys. Rev. Lett.},
	pages = {251601},
	primaryclass = {hep-th},
	title = {{Multiloop integrals in dimensional regularization made simple}},
	volume = {110},
	year = {2013}}

@article{Gehrmann:1999as,
	archiveprefix = {arXiv},
	author = {Gehrmann, T. and Remiddi, E.},
	doi = {10.1016/S0550-3213(00)00223-6},
	eprint = {hep-ph/9912329},
	journal = {Nucl. Phys. B},
	pages = {485--518},
	reportnumber = {TTP-99-49},
	title = {{Differential equations for two-loop four-point functions}},
	volume = {580},
	year = {2000}}

@article{Remiddi:1997ny,
	archiveprefix = {arXiv},
	author = {Remiddi, Ettore},
	doi = {10.1007/BF03185566},
	eprint = {hep-th/9711188},
	journal = {Nuovo Cim. A},
	pages = {1435--1452},
	reportnumber = {DFUB-97-15, DFUB 97-15},
	title = {{Differential equations for Feynman graph amplitudes}},
	volume = {110},
	year = {1997}}

@article{ChenSymbol,
	author = {K.~T.~Chen},
	journal = {Bull.\ Amer.\ Math.\ Soc.},
	pages = {831},
	title = {{Iterated path integrals}},
	volume = {83},
	year = {1977}}

@article{Bonisch:2024nru,
	archiveprefix = {arXiv},
	author = {B\"onisch, Kilian and Duhr, Claude and Maggio, Sara},
	eprint = {2404.04085},
	month = {4},
	primaryclass = {math.NT},
	reportnumber = {BONN-TH-06, MPIM-Bonn-2024},
	title = {{Some conjectures around magnetic modular forms}},
	year = {2024}}

@article{magnetic3,
	author = {Pasol, Vicentiu and Zudilin, Wadim},
	journal = {Nagoya Math. J.},
	pages = {849-864},
	title = {Magnetic (quasi-)modular forms},
	volume = {248},
	year = {2022}}

@article{magnetic1,
	author = {Broadhurst, David J. and Zudilin, Wadim},
	journal = {J. Austral. Math. Soc.},
	number = {1},
	pages = {9-25},
	title = {{A magnetic double integral}},
	volume = {107},
	year = {2019}}

@article{Duhr:2024uid,
	archiveprefix = {arXiv},
	author = {Duhr, Claude and Porkert, Franziska and Stawinski, Sven F.},
	doi = {10.1007/JHEP02(2025)014},
	eprint = {2412.02300},
	journal = {JHEP},
	pages = {014},
	primaryclass = {hep-th},
	reportnumber = {BONN-TH-2024-17},
	title = {{Canonical differential equations beyond genus one}},
	volume = {02},
	year = {2025}}

@article{Baune:2024biq,
	archiveprefix = {arXiv},
	author = {Baune, Konstantin and Broedel, Johannes and Im, Egor and Lisitsyn, Artyom and Zerbini, Federico},
	doi = {10.1088/1751-8121/ad8197},
	eprint = {2406.10051},
	journal = {J. Phys. A},
	number = {44},
	pages = {445202},
	primaryclass = {hep-th},
	title = {{Schottky\textendash{}Kronecker forms and hyperelliptic polylogarithms}},
	volume = {57},
	year = {2024}}

@article{DHoker:2025szl,
	archiveprefix = {arXiv},
	author = {D'Hoker, Eric and Enriquez, Benjamin and Schlotterer, Oliver and Zerbini, Federico},
	eprint = {2501.07640},
	month = {1},
	primaryclass = {hep-th},
	title = {{Relating flat connections and polylogarithms on higher genus Riemann surfaces}},
	year = {2025}}

@article{DHoker:2024ozn,
	archiveprefix = {arXiv},
	author = {D'Hoker, Eric and Schlotterer, Oliver},
	eprint = {2407.11476},
	month = {7},
	primaryclass = {hep-th},
	reportnumber = {UUITP--17/24},
	title = {{Fay identities for polylogarithms on higher-genus Riemann surfaces}},
	year = {2024}}

@inproceedings{ManinModular,
	address = {Boston},
	archiveprefix = {arXiv},
	author = {Manin, Y.~I.},
	booktitle = {Algebraic geometry and number theory},
	eprint = {math/0502576},
	pages = {565--597},
	publisher = {Birkh\"auser Boston},
	series = {Progr. Math.},
	title = {{Iterated integrals of modular forms and noncommutative modular symbols}},
	volume = {253},
	year = {2006}}

@article{Broedel:2014vla,
	archiveprefix = {arXiv},
	author = {Broedel, Johannes and Mafra, Carlos R. and Matthes, Nils and Schlotterer, Oliver},
	doi = {10.1007/JHEP07(2015)112},
	eprint = {1412.5535},
	journal = {JHEP},
	pages = {112},
	primaryclass = {hep-th},
	reportnumber = {AEI-2014-066, DAMTP-2014-95},
	title = {{Elliptic multiple zeta values and one-loop superstring amplitudes}},
	volume = {07},
	year = {2015}}

@article{LevinRacinet,
	author = {A. Levin and G. Racinet},
	eprint = {math/0703237},
	primaryclass = {math},
	title = {{Towards multiple elliptic polylogarithms}},
	year = {2007}}

@incollection{MR1265553,
	author = {Be\u{\i}linson, A. and Levin, A.},
	booktitle = {Motives ({S}eattle, {WA}, 1991)},
	doi = {10.1007/s00208-018-1645-4},
	mrclass = {11G05 (11G09 11G40 14H52 19F27)},
	mrnumber = {1265553},
	mrreviewer = {Jerzy Browkin},
	pages = {123--190},
	publisher = {Amer. Math. Soc., Providence, RI},
	series = {Proc. Sympos. Pure Math.},
	title = {The elliptic polylogarithm},
	url = {https://doi.org/10.1007/s00208-018-1645-4},
	volume = {55},
	year = {1994}}

@article{BrownLevin,
	archiveprefix = {arXiv},
	author = {Francis Brown and Andrey Levin},
	eprint = {1110.6917},
	primaryclass = {math.NT},
	title = {{Multiple Elliptic Polylogarithms}},
	year = {2011}}

@article{Laporta:2004rb,
	archiveprefix = {arXiv},
	author = {Laporta, S. and Remiddi, E.},
	doi = {10.1016/j.nuclphysb.2004.10.044},
	eprint = {hep-ph/0406160},
	journal = {Nucl. Phys. B},
	pages = {349--386},
	reportnumber = {CERN-PH-TH-2004-089},
	title = {{Analytic treatment of the two loop equal mass sunrise graph}},
	volume = {704},
	year = {2005}}

@article{Remiddi:1999ew,
	archiveprefix = {arXiv},
	author = {Remiddi, E. and Vermaseren, J. A. M.},
	doi = {10.1142/S0217751X00000367},
	eprint = {hep-ph/9905237},
	journal = {Int. J. Mod. Phys. A},
	pages = {725--754},
	reportnumber = {NIKHEF-99-005, TTP-99-08},
	title = {{Harmonic polylogarithms}},
	volume = {15},
	year = {2000}}

@article{Dlapa:2024cje,
	archiveprefix = {arXiv},
	author = {Dlapa, Christoph and K\"alin, Gregor and Liu, Zhengwen and Porto, Rafael A.},
	doi = {10.1103/PhysRevLett.132.221401},
	eprint = {2403.04853},
	journal = {Phys. Rev. Lett.},
	number = {22},
	pages = {221401},
	primaryclass = {hep-th},
	reportnumber = {DESY 24-029},
	title = {{Local in Time Conservative Binary Dynamics at Fourth Post-Minkowskian Order}},
	volume = {132},
	year = {2024}}

@article{Mishnyakov:2023wpd,
	archiveprefix = {arXiv},
	author = {Mishnyakov, Victor and Morozov, Alexei and Suprun, Pavel},
	doi = {10.1016/j.nuclphysb.2023.116245},
	eprint = {2303.08851},
	journal = {Nucl. Phys. B},
	pages = {116245},
	primaryclass = {hep-th},
	reportnumber = {MIPT/TH-09/23, FIAN/TH-07/23, ITEP/TH-10/23, IITP/TH-08/23},
	title = {{Position space equations for banana Feynman diagrams}},
	volume = {992},
	year = {2023}}

@article{Mishnyakov:2023sly,
	archiveprefix = {arXiv},
	author = {Mishnyakov, V. and Morozov, A. and Reva, M.},
	doi = {10.1016/j.nuclphysb.2024.116746},
	eprint = {2311.13524},
	journal = {Nucl. Phys. B},
	pages = {116746},
	primaryclass = {hep-th},
	reportnumber = {MIPT/TH-17/23 IITP/TH-16/23 ITEP/TH-22/23},
	title = {{On factorization hierarchy of equations for banana Feynman integrals}},
	volume = {1010},
	year = {2025}}

@article{Clingher2010LatticePK,
	author = {Adrian Clingher and Charles F. Doran},
	journal = {Advances in Mathematics},
	pages = {172-212},
	title = {Lattice polarized K3 surfaces and Siegel modular forms},
	url = {https://api.semanticscholar.org/CorpusID:7202956},
	volume = {231},
	year = {2010}}

@article{doranclingher,
	author = {Charles Doran and Adrian Clingher and J. Lewis and U. Whitcher},
	journal = {CRM-AMS proceedings and lecture notes},
	pages = {81-98},
	title = {Normal forms, K3 surface moduli, and modular parametrizations},
	volume = {47},
	year = {2009}}

@article{Gorges:2023zgv,
	archiveprefix = {arXiv},
	author = {G\"orges, Lennard and Nega, Christoph and Tancredi, Lorenzo and Wagner, Fabian J.},
	doi = {10.1007/JHEP07(2023)206},
	eprint = {2305.14090},
	journal = {JHEP},
	pages = {206},
	primaryclass = {hep-th},
	title = {{On a procedure to derive \ensuremath{\epsilon}-factorised differential equations beyond polylogarithms}},
	volume = {07},
	year = {2023}}

@article{Broedel:2019kmn,
	archiveprefix = {arXiv},
	author = {Broedel, Johannes and Duhr, Claude and Dulat, Falko and Marzucca, Robin and Penante, Brenda and Tancredi, Lorenzo},
	doi = {10.1007/JHEP09(2019)112},
	eprint = {1907.03787},
	journal = {JHEP},
	pages = {112},
	primaryclass = {hep-th},
	reportnumber = {CP3-19-34, CERN-TH-2019-105, HU-Mathematik-2019-04, HU-EP-19/20, SLAC-PUB-17453, HU-EP-19/20, SLAC-PUB-17453},
	title = {{An analytic solution for the equal-mass banana graph}},
	volume = {09},
	year = {2019}}

@inproceedings{Broedel:2018rwm,
	archiveprefix = {arXiv},
	author = {Broedel, Johannes and Duhr, Claude and Dulat, Falko and Penante, Brenda and Tancredi, Lorenzo},
	booktitle = {{KMPB Conference}: {Elliptic Integrals, Elliptic Functions and Modular Forms in Quantum Field Theory}},
	doi = {10.1007/978-3-030-04480-0_6},
	eprint = {1807.00842},
	pages = {107--131},
	primaryclass = {hep-th},
	reportnumber = {CERN-TH-2018-152},
	title = {{From modular forms to differential equations for Feynman integrals}},
	year = {2019}}

@article{Adams:2018yfj,
	archiveprefix = {arXiv},
	author = {Adams, Luise and Weinzierl, Stefan},
	doi = {10.1016/j.physletb.2018.04.002},
	eprint = {1802.05020},
	journal = {Phys. Lett. B},
	pages = {270--278},
	primaryclass = {hep-ph},
	reportnumber = {MITP-18-011},
	title = {{The $\varepsilon$-form of the differential equations for Feynman integrals in the elliptic case}},
	volume = {781},
	year = {2018}}

@article{Adams:2017ejb,
	archiveprefix = {arXiv},
	author = {Adams, Luise and Weinzierl, Stefan},
	doi = {10.4310/CNTP.2018.v12.n2.a1},
	eprint = {1704.08895},
	journal = {Commun. Num. Theor. Phys.},
	pages = {193--251},
	primaryclass = {hep-ph},
	title = {{Feynman integrals and iterated integrals of modular forms}},
	volume = {12},
	year = {2018}}

@article{Kreimer:2022fxm,
	archiveprefix = {arXiv},
	author = {Kreimer, Dirk},
	doi = {10.1007/s11005-023-01660-4},
	eprint = {2202.05490},
	journal = {Lett. Math. Phys.},
	number = {2},
	pages = {38},
	primaryclass = {hep-th},
	reportnumber = {MaPhy-AvH/2022-01},
	title = {{Bananas: multi-edge graphs and their Feynman integrals}},
	volume = {113},
	year = {2023}}

@article{doranclingher1,
	author = {Adrian Clingher and Charles Doran},
	doi = {10.1307/mmj/1187646999},
	journal = {Michigan Mathematical Journal},
	number = {2},
	pages = {355 -- 393},
	publisher = {University of Michigan, Department of Mathematics},
	title = {{Modular invariants for lattice polarized K3 surfaces}},
	url = {https://doi.org/10.1307/mmj/1187646999},
	volume = {55},
	year = {2007}}

@article{Almkvist3,
	archiveprefix = {arXiv},
	author = {Almkvist, G.},
	eprint = {math/0612215},
	title = {{Calabi-Yau differential equations of degree 2 and 3 and Yifan Yang's pullback}},
	year = {2006}}

@article{Maier,
	adsurl = {http://adsabs.harvard.edu/abs/2006math.....11041M},
	author = {{Maier}, R.~S.},
	eprint = {math/0611041},
	journal = {ArXiv Mathematics e-prints},
	month = nov,
	title = {{On Rationally Parametrized Modular Equations}},
	year = 2006}

@article{Primo:2017ipr,
	archiveprefix = {arXiv},
	author = {Primo, Amedeo and Tancredi, Lorenzo},
	doi = {10.1016/j.nuclphysb.2017.05.018},
	eprint = {1704.05465},
	journal = {Nucl. Phys. B},
	pages = {316--356},
	primaryclass = {hep-ph},
	reportnumber = {TTP17-021},
	title = {{Maximal cuts and differential equations for Feynman integrals. An application to the three-loop massive banana graph}},
	volume = {921},
	year = {2017}}

@article{Primo:2016ebd,
	archiveprefix = {arXiv},
	author = {Primo, Amedeo and Tancredi, Lorenzo},
	doi = {10.1016/j.nuclphysb.2016.12.021},
	eprint = {1610.08397},
	journal = {Nucl. Phys. B},
	pages = {94--116},
	primaryclass = {hep-ph},
	reportnumber = {TTP16-046},
	title = {{On the maximal cut of Feynman integrals and the solution of their differential equations}},
	volume = {916},
	year = {2017}}

@article{Bosma:2017ens,
	archiveprefix = {arXiv},
	author = {Bosma, Jorrit and Sogaard, Mads and Zhang, Yang},
	doi = {10.1007/JHEP08(2017)051},
	eprint = {1704.04255},
	journal = {JHEP},
	pages = {051},
	primaryclass = {hep-th},
	title = {{Maximal Cuts in Arbitrary Dimension}},
	volume = {08},
	year = {2017}}

@article{Broedel:2021zij,
	archiveprefix = {arXiv},
	author = {Broedel, Johannes and Duhr, Claude and Matthes, Nils},
	doi = {10.1007/JHEP02(2022)184},
	eprint = {2109.15251},
	journal = {JHEP},
	pages = {184},
	primaryclass = {hep-th},
	reportnumber = {BONN-TH-2021-08},
	title = {{Meromorphic modular forms and the three-loop equal-mass banana integral}},
	volume = {02},
	year = {2022}}

@article{10.1215/ijm/1258138437,
	author = {Yifan Yang and Noriko Yui},
	doi = {10.1215/ijm/1258138437},
	journal = {Illinois Journal of Mathematics},
	number = {2},
	pages = {667 -- 696},
	publisher = {Duke University Press},
	title = {{Differential equations satisfied by modular forms and $K3$ surfaces}},
	url = {https://doi.org/10.1215/ijm/1258138437},
	volume = {51},
	year = {2007}}

@article{doran,
	author = {Doran, Chuck},
	journal = {Comm. Math. Phys.},
	pages = {625--647},
	title = {{Picard--Fuchs Uniformization and Modularity of the Mirror Map}},
	volume = {212},
	year = {2000}}

@article{BognerCY,
	archiveprefix = {arXiv},
	author = {Bogner, Michael},
	eprint = {1304.5434},
	primaryclass = {math.AG},
	title = {{Algebraic characterization of differential operators of Calabi-Yau type}},
	year = {2013}}

@phdthesis{BognerThesis,
	author = {Bogner, Michael},
	school = {Johannes Gutenberg-Universit\"at Mainz},
	title = {{On differential operators of Calabi-Yau type}},
	year = {2012}}

@article{Bogner:2007mn,
	archiveprefix = {arXiv},
	author = {Bogner, Christian and Weinzierl, Stefan},
	doi = {10.1063/1.3106041},
	eprint = {0711.4863},
	journal = {J. Math. Phys.},
	pages = {042302},
	primaryclass = {hep-th},
	reportnumber = {MZ-TH-07-19},
	title = {{Periods and Feynman integrals}},
	volume = {50},
	year = {2009}}

@incollection{MR1852188,
	author = {Kontsevich, Maxim and Zagier, Don},
	booktitle = {Mathematics unlimited---2001 and beyond},
	mrclass = {11-02 (11F67 11G40 11G55)},
	mrnumber = {1852188},
	mrreviewer = {F. Beukers},
	pages = {771--808},
	publisher = {Springer, Berlin},
	title = {Periods},
	year = {2001}}

@inproceedings{Bourjaily:2022bwx,
	archiveprefix = {arXiv},
	author = {Bourjaily, Jacob L. and others},
	booktitle = {{Snowmass 2021}},
	eprint = {2203.07088},
	month = {3},
	primaryclass = {hep-ph},
	reportnumber = {BONN-TH-2022-05, UUITP-11/22, CERN-TH-2022-029, TUM-HEP-1391/22, HU-EP-22/08, MITP-22-022},
	title = {{Functions Beyond Multiple Polylogarithms for Precision Collider Physics}},
	year = {2022}}

@book{aomoto_theory_2011,
	author = {Aomoto, Kazuhiko and Kita, Michitake},
	publisher = {Springer Japan},
	series = {Springer {Monographs} in {Mathematics}},
	title = {Theory of {Hypergeometric} {Functions}},
	year = {2011}}

@book{yoshida_hypergeometric_1997,
	address = {Wiesbaden},
	author = {Yoshida, Masaaki},
	doi = {10.1007/978-3-322-90166-8},
	editor = {Diederich, Klas},
	isbn = {978-3-322-90168-2 978-3-322-90166-8},
	language = {en},
	publisher = {Vieweg+Teubner Verlag},
	series = {Aspects of {Mathematics}},
	title = {Hypergeometric {Functions}, {My} {Love}},
	url = {http://link.springer.com/10.1007/978-3-322-90166-8},
	urldate = {2023-01-06},
	volume = {32},
	year = {1997}}

@misc{matsumoto_relative_2019-1,
	author = {Matsumoto, Keiji},
	doi = {10.48550/arXiv.1804.00366},
	month = dec,
	publisher = {arXiv},
	title = {Relative twisted homology and cohomology groups associated with {Lauricella}'s \${F}\_D\$},
	url = {http://arxiv.org/abs/1804.00366},
	urldate = {2022-09-28},
	year = {2019}}

@article{Fontana:2023amt,
	archiveprefix = {arXiv},
	author = {Fontana, Gaia and Peraro, Tiziano},
	doi = {10.1007/JHEP08(2023)175},
	eprint = {2304.14336},
	journal = {JHEP},
	pages = {175},
	primaryclass = {hep-ph},
	reportnumber = {ZU-TH 19/23},
	title = {{Reduction to master integrals via intersection numbers and polynomial expansions}},
	volume = {08},
	year = {2023}}

@misc{Mpls1,
	author = {Goncharov, A. B.},
	doi = {10.48550/arXiv.math/0103059},
	month = may,
	note = {arXiv:math/0103059},
	publisher = {arXiv},
	title = {Multiple polylogarithms and mixed {Tate} motives},
	url = {http://arxiv.org/abs/math/0103059},
	urldate = {2023-08-03},
	year = {2001}}

@article{Bloch:2014qca,
	archiveprefix = {arXiv},
	author = {Bloch, Spencer and Kerr, Matt and Vanhove, Pierre},
	doi = {10.1112/S0010437X15007472},
	eprint = {1406.2664},
	journal = {Compos. Math.},
	number = {12},
	pages = {2329--2375},
	primaryclass = {hep-th},
	reportnumber = {IPHT-T-14-015, IHES-P-14-06},
	title = {{A Feynman integral via higher normal functions}},
	volume = {151},
	year = {2015}}

@article{MR3780269,
	archiveprefix = {arXiv},
	author = {Bloch, Spencer and Kerr, Matt and Vanhove, Pierre},
	doi = {10.4310/ATMP.2017.v21.n6.a1},
	eprint = {1601.08181},
	fjournal = {Advances in Theoretical and Mathematical Physics},
	issn = {1095-0761},
	journal = {Adv. Theor. Math. Phys.},
	mrclass = {14J33 (14D07 33E30 34C14 81S40)},
	mrnumber = {3780269},
	mrreviewer = {Alan Matthew Thompson},
	number = {6},
	pages = {1373--1454},
	primaryclass = {hep-th},
	title = {Local mirror symmetry and the sunset {F}eynman integral},
	url = {https://doi.org/10.4310/ATMP.2017.v21.n6.a1},
	volume = {21},
	year = {2017}}

@article{Klemm:2019dbm,
	archiveprefix = {arXiv},
	author = {Klemm, Albrecht and Nega, Christoph and Safari, Reza},
	doi = {10.1007/JHEP04(2020)088},
	eprint = {1912.06201},
	journal = {JHEP},
	pages = {088},
	primaryclass = {hep-th},
	reportnumber = {BONN-TH-2019-07},
	title = {{The $l$-loop Banana Amplitude from GKZ Systems and relative Calabi-Yau Periods}},
	volume = {04},
	year = {2020}}

@article{Bonisch:2020qmm,
	archiveprefix = {arXiv},
	author = {B\"onisch, Kilian and Fischbach, Fabian and Klemm, Albrecht and Nega, Christoph and Safari, Reza},
	doi = {10.1007/JHEP05(2021)066},
	eprint = {2008.10574},
	journal = {JHEP},
	pages = {066},
	primaryclass = {hep-th},
	reportnumber = {BONN-TH-2020-06},
	title = {{Analytic structure of all loop banana integrals}},
	volume = {05},
	year = {2021}}

@article{Bonisch:2021yfw,
	archiveprefix = {arXiv},
	author = {B\"onisch, Kilian and Duhr, Claude and Fischbach, Fabian and Klemm, Albrecht and Nega, Christoph},
	doi = {10.1007/JHEP09(2022)156},
	eprint = {2108.05310},
	journal = {JHEP},
	pages = {156},
	primaryclass = {hep-th},
	reportnumber = {BONN-TH-2021-05},
	title = {{Feynman integrals in dimensional regularization and extensions of Calabi-Yau motives}},
	volume = {09},
	year = {2022}}

@article{Pogel:2022ken,
	archiveprefix = {arXiv},
	author = {P\"ogel, Sebastian and Wang, Xing and Weinzierl, Stefan},
	doi = {10.1103/PhysRevLett.130.101601},
	eprint = {2211.04292},
	journal = {Phys. Rev. Lett.},
	number = {10},
	pages = {101601},
	primaryclass = {hep-th},
	reportnumber = {MITP/22-094, TUM-HEP-1431/22},
	title = {{Taming Calabi-Yau Feynman Integrals: The Four-Loop Equal-Mass Banana Integral}},
	volume = {130},
	year = {2023}}

@article{Pogel:2022vat,
	archiveprefix = {arXiv},
	author = {P\"ogel, Sebastian and Wang, Xing and Weinzierl, Stefan},
	doi = {10.1007/JHEP04(2023)117},
	eprint = {2212.08908},
	journal = {JHEP},
	pages = {117},
	primaryclass = {hep-th},
	title = {{Bananas of equal mass: any loop, any order in the dimensional regularisation parameter}},
	volume = {04},
	year = {2023}}

@article{Pogel:2022yat,
	archiveprefix = {arXiv},
	author = {P\"ogel, Sebastian and Wang, Xing and Weinzierl, Stefan},
	doi = {10.1007/JHEP09(2022)062},
	eprint = {2207.12893},
	journal = {JHEP},
	pages = {062},
	primaryclass = {hep-th},
	title = {{The three-loop equal-mass banana integral in \ensuremath{\varepsilon}-factorised form with meromorphic modular forms}},
	volume = {09},
	year = {2022}}

@article{Mizera:2017rqa,
	archiveprefix = {arXiv},
	author = {Mizera, Sebastian},
	doi = {10.1103/PhysRevLett.120.141602},
	eprint = {1711.00469},
	journal = {Phys. Rev. Lett.},
	number = {14},
	pages = {141602},
	primaryclass = {hep-th},
	title = {{Scattering Amplitudes from Intersection Theory}},
	volume = {120},
	year = {2018}}

@article{Mastrolia:2018uzb,
	archiveprefix = {arXiv},
	author = {Mastrolia, Pierpaolo and Mizera, Sebastian},
	doi = {10.1007/JHEP02(2019)139},
	eprint = {1810.03818},
	journal = {JHEP},
	pages = {139},
	primaryclass = {hep-th},
	title = {{Feynman Integrals and Intersection Theory}},
	volume = {02},
	year = {2019}}

@article{Frellesvig:2019uqt,
	archiveprefix = {arXiv},
	author = {Frellesvig, Hjalte and Gasparotto, Federico and Mandal, Manoj K. and Mastrolia, Pierpaolo and Mattiazzi, Luca and Mizera, Sebastian},
	doi = {10.1103/PhysRevLett.123.201602},
	eprint = {1907.02000},
	journal = {Phys. Rev. Lett.},
	number = {20},
	pages = {201602},
	primaryclass = {hep-th},
	title = {{Vector Space of Feynman Integrals and Multivariate Intersection Numbers}},
	volume = {123},
	year = {2019}}

@article{Frellesvig:2019kgj,
	archiveprefix = {arXiv},
	author = {Frellesvig, Hjalte and Gasparotto, Federico and Laporta, Stefano and Mandal, Manoj K. and Mastrolia, Pierpaolo and Mattiazzi, Luca and Mizera, Sebastian},
	doi = {10.1007/JHEP05(2019)153},
	eprint = {1901.11510},
	journal = {JHEP},
	pages = {153},
	primaryclass = {hep-ph},
	title = {{Decomposition of Feynman Integrals on the Maximal Cut by Intersection Numbers}},
	volume = {05},
	year = {2019}}

@article{Frellesvig:2020qot,
	archiveprefix = {arXiv},
	author = {Frellesvig, Hjalte and Gasparotto, Federico and Laporta, Stefano and Mandal, Manoj K. and Mastrolia, Pierpaolo and Mattiazzi, Luca and Mizera, Sebastian},
	doi = {10.1007/JHEP03(2021)027},
	eprint = {2008.04823},
	journal = {JHEP},
	pages = {027},
	primaryclass = {hep-th},
	title = {{Decomposition of Feynman Integrals by Multivariate Intersection Numbers}},
	volume = {03},
	year = {2021}}

@article{Caron-Huot:2021xqj,
	archiveprefix = {arXiv},
	author = {Caron-Huot, Simon and Pokraka, Andrzej},
	doi = {10.1007/JHEP12(2021)045},
	eprint = {2104.06898},
	journal = {JHEP},
	pages = {045},
	primaryclass = {hep-th},
	title = {{Duals of Feynman integrals. Part I. Differential equations}},
	volume = {12},
	year = {2021}}

@article{Caron-Huot:2021iev,
	archiveprefix = {arXiv},
	author = {Caron-Huot, Simon and Pokraka, Andrzej},
	doi = {10.1007/JHEP04(2022)078},
	eprint = {2112.00055},
	journal = {JHEP},
	pages = {078},
	primaryclass = {hep-th},
	title = {{Duals of Feynman Integrals. Part II. Generalized unitarity}},
	volume = {04},
	year = {2022}}

@article{Chestnov:2022xsy,
	archiveprefix = {arXiv},
	author = {Chestnov, Vsevolod and Frellesvig, Hjalte and Gasparotto, Federico and Mandal, Manoj K. and Mastrolia, Pierpaolo},
	doi = {10.1007/JHEP06(2023)131},
	eprint = {2209.01997},
	journal = {JHEP},
	pages = {131},
	primaryclass = {hep-th},
	title = {{Intersection numbers from higher-order partial differential equations}},
	volume = {06},
	year = {2023}}

@article{Duhr:2025ppd,
	archiveprefix = {arXiv},
	author = {Duhr, Claude and Maggio, Sara},
	doi = {10.1007/JHEP06(2025)250},
	eprint = {2502.15326},
	journal = {JHEP},
	pages = {250},
	primaryclass = {hep-th},
	reportnumber = {BONN-TH-2025-05},
	title = {{Feynman integrals, elliptic integrals and two-parameter K3 surfaces}},
	volume = {06},
	year = {2025}}

@article{Lee:2013mka,
	archiveprefix = {arXiv},
	author = {Lee, Roman N.},
	doi = {10.1088/1742-6596/523/1/012059},
	editor = {Wang, Jianxiong},
	eprint = {1310.1145},
	journal = {J. Phys. Conf. Ser.},
	pages = {012059},
	primaryclass = {hep-ph},
	title = {{LiteRed 1.4: a powerful tool for reduction of multiloop integrals}},
	volume = {523},
	year = {2014}}

@article{Duhr:2025tdf,
	archiveprefix = {arXiv},
	author = {Duhr, Claude},
	doi = {10.1007/JHEP08(2025)218},
	eprint = {2502.15325},
	journal = {JHEP},
	pages = {218},
	primaryclass = {hep-th},
	reportnumber = {BONN-TH-2025-04},
	title = {{Modular forms for three-loop banana integrals}},
	volume = {08},
	year = {2025}}

@article{Duhr:2025kkq,
	archiveprefix = {arXiv},
	author = {Duhr, Claude and Maggio, Sara and Porkert, Franziska and Semper, Cathrin and Stawinski, Sven F.},
	eprint = {2507.23061},
	month = {7},
	primaryclass = {hep-th},
	reportnumber = {BONN-TH-2025-24},
	title = {{Three-loop banana integrals with four unequal masses}},
	year = {2025}}

@article{Pogel:2025bca,
	archiveprefix = {arXiv},
	author = {P{\"o}gel, Sebastian and Teschke, Toni and Wang, Xing and Weinzierl, Stefan},
	eprint = {2507.23594},
	month = {7},
	primaryclass = {hep-th},
	title = {{The unequal-mass three-loop banana integral}},
	year = {2025}}

@phdthesis{Mizera:2019gea,
	archiveprefix = {arXiv},
	author = {Mizera, Sebastian},
	doi = {10.1007/978-3-030-53010-5},
	eprint = {1906.02099},
	primaryclass = {hep-th},
	school = {Princeton, Inst. Advanced Study},
	title = {{Aspects of Scattering Amplitudes and Moduli Space Localization}},
	year = {2020}}

@article{Weinzierl:2020xyy,
	archiveprefix = {arXiv},
	author = {Weinzierl, Stefan},
	doi = {10.1063/5.0054292},
	eprint = {2002.01930},
	journal = {J. Math. Phys.},
	number = {7},
	pages = {072301},
	primaryclass = {math-ph},
	title = {{On the computation of intersection numbers for twisted cocycles}},
	volume = {62},
	year = {2021}}

@article{Weinzierl:2020nhw,
	archiveprefix = {arXiv},
	author = {Weinzierl, Stefan},
	doi = {10.1016/j.physletb.2020.135449},
	eprint = {2003.05839},
	journal = {Phys. Lett. B},
	pages = {135449},
	primaryclass = {hep-th},
	reportnumber = {MITP/20-013},
	title = {{Correlation functions on the lattice and twisted cocycles}},
	volume = {805},
	year = {2020}}

@article{Cacciatori:2021nli,
	archiveprefix = {arXiv},
	author = {Cacciatori, Sergio Luigi and Conti, Maria and Trevisan, Simone},
	doi = {10.3390/universe7090328},
	eprint = {2107.14721},
	journal = {Universe},
	number = {9},
	pages = {328},
	primaryclass = {hep-th},
	title = {{Co-Homology of Differential Forms and Feynman Diagrams}},
	volume = {7},
	year = {2021}}

@article{Brunello:2023rpq,
	archiveprefix = {arXiv},
	author = {Brunello, Giacomo and Chestnov, Vsevolod and Crisanti, Giulio and Frellesvig, Hjalte and Mandal, Manoj K. and Mastrolia, Pierpaolo},
	doi = {10.1007/JHEP09(2024)015},
	eprint = {2401.01897},
	journal = {JHEP},
	pages = {015},
	primaryclass = {hep-th},
	title = {{Intersection numbers, polynomial division and relative cohomology}},
	volume = {09},
	year = {2024}}

@article{Brunello:2024tqf,
	archiveprefix = {arXiv},
	author = {Brunello, Giacomo and Chestnov, Vsevolod and Mastrolia, Pierpaolo},
	doi = {10.1007/JHEP07(2025)045},
	eprint = {2408.16668},
	journal = {JHEP},
	pages = {045},
	primaryclass = {hep-th},
	title = {{Intersection numbers from companion tensor algebra}},
	volume = {07},
	year = {2025}}

@article{Crisanti:2024onv,
	archiveprefix = {arXiv},
	author = {Crisanti, Giulio and Smith, Sid},
	doi = {10.1007/JHEP09(2024)018},
	eprint = {2405.18178},
	journal = {JHEP},
	pages = {018},
	primaryclass = {hep-th},
	reportnumber = {MPP-2024-104},
	title = {{Feynman integral reductions by intersection theory with orthogonal bases and closed formulae}},
	volume = {09},
	year = {2024}}

@article{Lu:2024dsb,
	archiveprefix = {arXiv},
	author = {Lu, Mingming and Wang, Ziwen and Yang, Li Lin},
	doi = {10.1007/JHEP05(2025)158},
	eprint = {2411.05226},
	journal = {JHEP},
	pages = {158},
	primaryclass = {hep-th},
	title = {{Intersection theory, relative cohomology and the Feynman parametrization}},
	volume = {05},
	year = {2025}}

@article{Duhr:2024xsy,
	archiveprefix = {arXiv},
	author = {Duhr, Claude and Porkert, Franziska and Semper, Cathrin and Stawinski, Sven F.},
	doi = {10.1007/JHEP03(2025)053},
	eprint = {2408.04904},
	journal = {JHEP},
	pages = {053},
	primaryclass = {hep-th},
	reportnumber = {BONN-TH-2024-11},
	title = {{Self-duality from twisted cohomology}},
	volume = {03},
	year = {2025}}

@article{Liu:2022chg,
	archiveprefix = {arXiv},
	author = {Liu, Xiao and Ma, Yan-Qing},
	doi = {10.1016/j.cpc.2022.108565},
	eprint = {2201.11669},
	journal = {Comput. Phys. Commun.},
	pages = {108565},
	primaryclass = {hep-ph},
	title = {{AMFlow: A Mathematica package for Feynman integrals computation via auxiliary mass flow}},
	volume = {283},
	year = {2023}}

\end{document}